\documentclass[12pt]{article}
\usepackage{amsmath}
\usepackage{graphicx}
\usepackage{enumerate}
\usepackage[numbers]{natbib}
\usepackage{url}
\usepackage{color}
\usepackage{float}
\usepackage[utf8]{inputenc}
\usepackage{amsthm} 
\usepackage{amssymb}
\usepackage{multirow}
\usepackage{latexsym}
\usepackage{enumitem}

\def\d{{\mathrm d}}

\newtheorem{remark}{Remark}

\newcommand{\blind}{1}

\usepackage{float}
\def\d{{\mathrm d}} 

\usepackage[plain,noend]{algorithm2e}
\RestyleAlgo{ruled,vlined}  
\SetKwComment{tcc}{\* }{}    
\makeatletter
\renewcommand{\algocf@captiontext}[2]{#1\algocf@typo. \AlCapFnt{}#2} 
\def\@algocf@capt@plain{top}
\renewcommand{\algocf@makecaption}[2]{%
	\addtolength{\hsize}{\algomargin}%
	\sbox\@tempboxa{\algocf@captiontext{#1}{#2}}%
	\ifdim\wd\@tempboxa >\hsize
	\hskip .5\algomargin%
	\parbox[t]{\hsize}{\algocf@captiontext{#1}{#2}}
	\else%
	\global\@minipagefalse%
	\hbox to\hsize{\box\@tempboxa}
	\fi%
	\addtolength{\hsize}{-\algomargin}%
}
\makeatother

\begin{document}


\def\spacingset#1{\renewcommand{\baselinestretch}%
	{#1}\small\normalsize} \spacingset{1}


\if1\blind
{
	\title{\bf Monitoring  a developing pandemic with available data 
	}
	\author{
		Mar{\'\i}a Luz G\'amiz\thanks{Corresponding author: mgamiz@ugr.es} \\ Department of Statistics and O.R., University of Granada, Spain\\ \\
		Enno Mammen\\ Institute of Applied Mathematics, Heidelberg University, Germany\\ \\
		Mar{\'\i}a Dolores Mart{\'\i}nez-Miranda
		\\ Department of Statistics and O.R., University of Granada, Spain\\ \\
		Jens Perch Nielsen \\
		Bayes Business School, City, University of London, London, UK \\ \\
		Michael Scholz \\Department of Economics, University of Graz, Austria \\ \\
		Germán Ernesto Silva-Gómez \\ Department of Statistics and O.R., University of Granada, Spain
	}
	\maketitle
} \fi

\if0\blind
{
	\bigskip
	\bigskip
	\bigskip
	\begin{center}
		{\LARGE\bf  Monitoring a developing pandemic with available data}
	\end{center}
	\medskip
} \fi

\vspace{-1cm}
\begin{abstract}
This paper addresses statistical modelling and forecasting of key indicators describing the severity of a developing pandemic, using routinely reported daily counts of infections, hospitalizations, deaths (both in and out of hospital), and recoveries. These observed counts constitute what we term  ``available data''.
Because such data are typically incomplete or inconsistently reported, we address several novel missing data challenges arising in this context and propose statistically rigorous solutions that enable inference based solely on the available information. The model is formulated dynamically, explicitly incorporating calendar effects to capture systematic temporal variations in the progression of the pandemic. 
The proposed framework is illustrated using data from France collected during the COVID-19 pandemic. Our approach also establishes a new benchmark for integrating prior information from domain experts directly into the modelling process, thereby enabling a potential new division of labour between statistical estimation and epidemiological knowledge from external experts.

\end{abstract}

\noindent%
\vfill

\newpage
\spacingset{1.5} 

\section{Introduction}\label{sec:intro}
Let $N_1$, $N_2$, $N_3$ and $N_4$ be four time series representing respectively, the daily number of positive tests for infection ($N_1$), hospital admissions ($N_2$),  hospital discharges ($N_3$), and hospital deaths ($N_4$).  Let $F$ denote a fifth time series measuring the ratio of deaths occurring outside hospital to those occurring within hospital.

A joint model for the stochastic evolution of these five processes can provide a natural benchmark for monitoring the progression of a pandemic. An advantage of this formulation is that the four counting processes $N_1$, $N_2$, $N_3$, $N_4$  are typically available from routine surveillance data, even at early stages of an outbreak in most countries or regions.
The fifth process $F$ is likewise available, since deaths are typically recorded with high accuracy.

The main challenge of this approach lies in characterizing the statistical dependence between the counting processes, for example between $N_1$ and $N_2$.A fundamental difficulty is that the data lack individual-level linkage: there are no records identifying who was infected when, or when the same individuals entered or left hospital. Only the aggregated counts from the four processes  $N_1$, $N_2$, $N_3$, $N_4$ are observed.

This missing link estimation problem has been investigated to some extent in probabilistic queueing theory  see for example \cite{Goldenshluger:Koops:19}. While the work of \cite{Goldenshluger:Koops:19} studied nonparametric estimation of service-time distributions in infinite-server queues with exogenous, nonstationary Poisson arrivals, the present work extends this framework to a more realistic and complex setting in which  both arrivals and departures are endogenous and time-dependent. Specifically,  daily hospital admissions and deaths or recoveries are modelled by defining the intensity rate or survival hazard as a functional of past infections and admissions through a time-varying kernel. This formulation accommodates calendar effects, feedback between inflows and outflows, and nonstationary survival dynamics, features that are absent in the purely exogenous model of  \cite{Goldenshluger:Koops:19}. The resulting stochastic, dynamic system generalizes their deconvolution problem and yields a flexible framework for estimating length-of-stay distributions, providing a practical basis for pandemic monitoring.

Another complication, which to our knowledge, has not been addressed in the queuing literature, is that $N_1(t)$ does not represent the true number of infections on day $t$, but rather the number of individuals who test positive, a fraction of the number of new infections that constitute the actual exposure driving subsequent transmissions and hospitalizations. In other words, the observed exposure underreports the true exposure. Underreporting is a well recognized challenge in the analysis of epidemic and other real-world time series. It can distort key inferences, such as the strength of dependence  \cite{Wei:etal:24}. Correction methods of varying complexity have been proposed in the literature (see for example \cite{Bracher:Held:2021}).

In contrast, the proposed approach does not require correction for underreported exposure: it models the observed $N_1(t)$ data directly while still providing accurate predictions of hospitalizations, deaths and discharges.\\
Missing-link survival analysis was introduced in \cite{Gamiz:etal:22} to estimate individual-level length-of-stay from aggregated counts when the connection between admissions and discharges is unobserved. Building on this framework, the present approach extends the missing-link paradigm by explicitly incorporating calendar effects, allowing for nonstationary admission and exit processes, and improving estimation and forecasting of hospital stay durations in pandemic settings.

On the other hand, the term ``exposure of low quality'' has been introduced in \cite{Gamiz:etal:25},  to describe the changing nature of the exposure information provided by $N_1$ when forecasting future values of $N_1$ and $N_2$. This paper extends that methodology to model hospital survival processes, including in-hospital mortality and discharge dynamics, thereby illustrating the full potential of the methodology introduced in \cite{Gamiz:etal:25}. 
These methods are particularly useful for policymakers in the early stages of a pandemic, when crucial decisions must be made with limited data. As noted by \cite{Britton:Tomba:19}, estimates based on early outbreak data are often biased due to right-censoring, underreporting, and changes in case ascertainment over time. More specifically, they conclude that early epidemic estimates, including the reproduction number $R$ (\cite{Fraser:07}), are particularly prone to bias, stressing the need for rigorous statistical methods to obtain reliable and accurate estimates.

While numerous methods exist for estimating pandemic dynamics (see, for example, \cite{Slater:etal:23}, \cite{Mahsin:etal:22}, \cite{Sparapani:etal:20}, \cite{Quick:etal:21}) none provide a simple benchmark approach based solely on the readily available counting processes $N_1$, $N_2$, $N_3$, $N_4$ and $F$. \cite{Slater:etal:23} estimate cumulative incidence and infection fatality rates directly from continuous serological data; \cite{Mahsin:etal:22} rely on detailed spatial and individual-level data; \cite{Sparapani:etal:20} propose nonparametric recurrent-event models applied to individual-level hospital admission data; \cite{Quick:etal:21} develop a multilevel regression framework to jointly estimate incidence, prevalence, and reproductive numbers using confirmed infections, serological surveys, and testing data.

The purpose of this paper is to introduce a benchmark procedure for epidemic monitoring and forecasting that can be implemented directly from routinely collected data. Motivated by experience during the Covid-19 pandemic, we assume that $N_1$, $N_2$, $N_3$, $N_4$ and $F$ are typically observed, possibly supplemented by prior information on the reproduction number \citep{Fraser:07} and on the ratio of out-of-hospital to in-hospital deaths. The framework provides a coherent means of incorporating such prior information and expert judgment, which may evolve as the epidemic progresses. During the Covid-19 pandemic, the reproduction number ($R$) was widely reported and interpreted as a measure of transmission intensity; the proposed method formally accommodates this and related quantities, including expert assessments of mortality across settings, and is consistent with regression-based estimation of $R$ as in \cite{Quick:etal:21}. The procedure is based on transparent assumptions and interpretable quantities, thereby facilitating statistical communication and enabling the structured inclusion of expert information within a simple inferential framework.

The main methodological contributions are as follows.
\begin{enumerate}[label=(\roman*)]
	\item Building on earlier work in \cite{Gamiz:etal:25} that models $N_1$ and $N_2$ as self-exciting point processes, we extend the framework to the subsequent processes $N_3$ and $N_4$, which are modelled within the Aalen multiplicative intensity framework.
	\item We propose an iterative estimation algorithm for the corresponding two-dimensional survival hazards and provide theoretical justification for the resulting estimators.
	\item We develop a bootstrap procedure to quantify the uncertainty of both estimators and forecasts.
	\item We introduce a principled approach to incorporating auxiliary information or expert knowledge through two interpretable quantities, $C_1$ and $C_2$, enabling flexible and transparent adjustment of forecasts.
	\item Finally, we present a case study based on Covid-19 data from France, illustrating the joint modelling of $N_1$--$N_4$ and the integration of expert information derived from published reproduction numbers and assessments of in-hospital and out-of-hospital mortality.
\end{enumerate}

\noindent The organisation of the paper is as follows:\\
Section~\ref{sec:monitoring} introduces the transition model for monitoring the progression of a developing pandemic, with its formal formulation presented in Section~\ref{sec:model}. Section~\ref{sec:estim} describes the estimation procedure, and Section~\ref{sec:forecast} outlines the fundamentals of forecasting. In Section~\ref{sec:application}, the methodology is illustrated through an application to the Covid-19 pandemic in France. Section~\ref{sec:conclusion} concludes the paper.

The appendices include a finite-sample simulation study demonstrating that our proposed methodology effectively addresses challenges arising from missing-link survival analysis, the algorithm presented in stepwise form, the underlying theoretical results, and additional technical details. The code used to reproduce the results will be made available on CRAN as the R package {\tt Pandemics} (see \cite{Gamiz:etal:24}).

\section{General considerations when monitoring and forecasting in a dynamic environment}\label{sec:monitoring} 
\subsection{Modelling the dynamics of a pandemic}\label{sec:full_system} 
Our dynamic modelling is broadly speaking based on two types of transitions. One type where the number of individuals are well defined and a follow-up type survival analysis is possible which have been discussed in  Section \ref{sec:model2}. Another type of transition, in which the number of individuals are biased by dynamic definitions and underreporting, is fully addressed in  \cite{Gamiz:etal:25}.

\begin{figure}[ht]
	\centering
	\includegraphics[width=10cm]{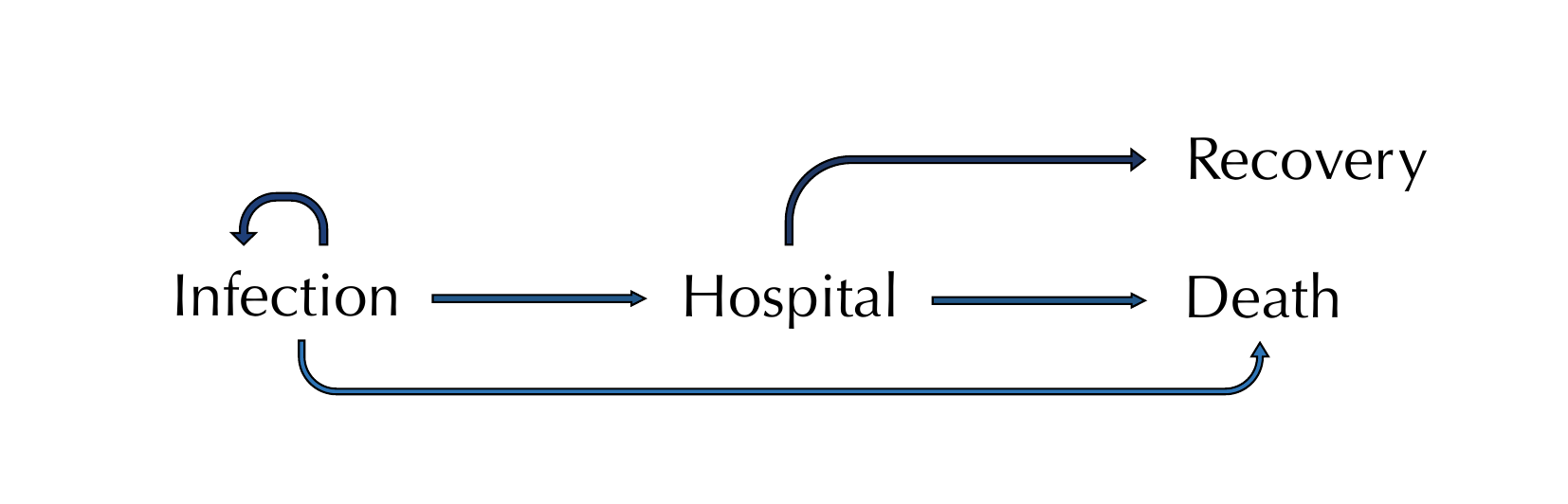}
	\caption{Transition diagram.} \label{fig:diagram}
\end{figure}

In this section we describe the full model chosen in the Covid-19 case of France. 
Our model can be illustrated via the stages represented in Figure \ref{fig:diagram}.  

In the early phase of a pandemic, only a few individuals are infected, and the rate of further infections increases with the number already infected, generating a feedback mechanism from the infected stage to itself. The definition of infection may change over time; a practical formulation is to define infected individuals as those with a confirmed positive test. In the initial stages, limited testing capacity implies that a positive test represents a higher fraction of total infections than at later stages when testing becomes widespread.

The transition rate from infection to hospitalization may vary with both calendar time  and duration in the infected stage, reflecting evolving clinical practices, testing policies, or seasonal effects. Similarly, the transitions from hospitalization to death or recovery may depend on the same arguments. While the definition of death itself is stable, the classification of a death as pandemic-related may vary over time. Figure~\ref{fig:diagram} illustrates the two possible outcomes following hospitalization—transition to death or recovery—with all transition intensities modelled as two-dimensional functions. The dependence on calendar time introduces the temporal dynamics of interest.

\subsection{ Transitions from hospitalized to dead or from hospitalized to recovered}\label{sec:outcome} 

Standard survival analysis cannot be directly applied to these transitions using the available data, which comprise daily counts of hospitalized individuals, deaths in hospital, and recoveries in hospital, but do not record the individual-level link between these events. Consequently, the duration until death or recovery for a given hospitalization is unobserved. This issue is addressed by introducing novel survival techniques for handling such missing information, as described in Sections \ref{sec:model} and \ref{sec:estim}.

\subsection{ Relationship between dying outside hospital and dying in hospital}\label{sec:deaths}
Available data suffice to model transitions from infected to infected, from infected to hospitalized, and from hospitalized to recovery or death. Total deaths in the population, corresponding to transitions from infected to death, require additional data on daily deaths outside the hospital. Let $N_{\rm in}(t)$ and $N_{\rm out}(t)$ denote the number of deaths in and outside hospitals at time $t>0$, and define the dynamic ratio $F(t)={N_{\rm out}(t)}/{N_{\rm in}(t)}$.

For $t\in(0,T)$, $F(t)$ is estimated as the ratio of two nonparametric regression estimators: a smoothed regression of $N_{\rm out}(t)$ over time in the numerator, and a smoothed regression of $N_{\rm in}(t)$ over time in the denominator, yielding $\widehat{F}(t)$. 

\section{Model formulation}\label{sec:model}
\subsection{Modelling infection feedback loops and transitions from infection to hospitalization}\label{sec:model1}

Using Hawkes processes (\cite{Raad:etal:20}, \cite{Mammen:Muller:23}), the transitions from $N_1$ to $N_1$ and from $N_1$ to $N_2$ are studied in  \cite{Gamiz:etal:25} with a new definition of ``exposure of low quality''. In the following, we summarize the procedure detailed in \cite{Gamiz:etal:25}.\\
\noindent We observe $n$ individuals over $(0,T]$, and let $N_1(t)$ and $N_2(t)$ denote, respectively, the cumulative numbers of infections and hospitalizations up to time $t \ $ ($0 < t \leq T$).
For $k$ large enough, $N_j(t)=\sum_{i=1}^k \left[N_j(t i/k) - N_j(t (i-1)/k)\right] \ $ ($j=1,2$), which, following usual notation in counting process theory, we also write as $N_j(t)=\int_0^t N_j(\d s)$. Thus, $N_1(\d s)$ can be interpreted as the number of new infections in the interval $(s, s+\d s]$, and $N_2(\d s)$ can be interpreted as the number of new hospitalized in the interval $(s, s+\d s]$. \\
To describe the  dynamics of the processes  $N_j(t)$ ($j=1,2$), let ${\mathcal F}_1(t)$ be the $\sigma$-field generated by $\{N_1(s): s<t\}$. Define the intensity
\begin{equation}\label{eq:lambda}
	\lambda_j(t) = \lim_{\d t\to 0}\frac{E[N_j(t+\d t)-N_j(t)\mid{\mathcal F}_1(t)]}{\d t} \quad (j=1,2).
\end{equation}
On the one hand, for the infection process $N_1(t)$, we model $\lambda_1(t)$ as a Hawkes process, that is
\begin{equation}\label{eq:haw1_reduced}
	\lambda_1(t)\d t = \Big(\int_0^{t^-}\mu_1(t/T,t-s)N_1(\d s)\Big)\d t + n\rho_1(t)\d t,
\end{equation}
where $\mu_1$ models the infection-to-infection transmission, and $\rho_1$ accounts for early or external cases but vanishes after an initial period.
On the other hand, for the hospitalization process $N_2(t)$, conditional on the infection history, the hospitalization intensity is
\begin{equation}\label{eq:haw2_reduced}
	\lambda_2(t)\d t = \Big(\int_0^{t^-}\mu_2(t/T,t-s)N_1(\d s)\Big)\d t + n\rho_2(t)\d t,
\end{equation}
where $\mu_2$ captures the delay from infection to hospitalization, and $\rho_2$ vanishes after the early phase. In the case of the hospitalizations, the model is not a Hawkes-type process. For $N_2(t)$ to be modelled as a Hawkes process,  in equation \eqref{eq:lambda} for $\lambda_2(t)$,  $\mathcal{F}_1(t)$  would need to be replaced by $\{(N_1(s),N_2(s)): s<t\}$.

Based on our motivating example of Covid-19, we conjecture that model \eqref{eq:haw2_reduced}, together with the history of the infection process $N_1$, is sufficient to predict future hospitalizations (additional methodological details are provided in \citep{Gamiz:etal:25}).\\

Our main interest is estimation of $\mu_1$ and $\mu_2$, which describe the transmission and progression dynamics. The nuisance terms $\rho_1$ and $\rho_2$ ensure flexibility at the start of the epidemic but are negligible in later periods.
\subsection{Modelling transitions from hospitalization to death or recovery}\label{sec:model2}
The hospitalized persons can leave the hospital due to death or recovery, whichever occurs first.  In this section we concentrate  on generalising the missing link survival analysis of \cite{Gamiz:etal:22} to the dynamic and multivariate situation needed to investigate the transitions from $N_2$ to $N_3$ and from $N_2$ to $N_4$. \\
Let us denote $N(t)$ the total number of patients that leave hospital in $(0,t]$, then $N(t)=N_3(t)+N_4(t)$, where as in Section \ref{sec:intro}, $N_3(t)$ counts the number of patients that receive medical discharge and $N_4(t)$ the number of patients that die in hospital in the interval $(0,t]$.  Using the notation of Section \ref{sec:deaths}, we also denote $N_4(t)=N_{\rm in}(t)$. Then $N(\d t)$ is the number of persons that leave the hospital (due to death or recovery) in the interval $(t, t+\d t]$, and $N(t)=\int_0^t N(\d s) \ $ ($0< t \leq T$). 
Furthermore, we write $N(\d t,\d s)$ for the number of persons who entered the hospital in $(s,s+ \d s]$ and leave in  $(t,t+\d t]$ due to any cause  ($s<t \leq T $). \\
Let $\lambda$ denote the intensity function of $N$, that is, 
\[
\lambda(t)=\lim_{\d t \rightarrow 0} \frac{\text{pr}(N(t+\d t)-N(t)\geq 1 \mid {\cal F}(t))}{ \d t} \qquad  (0 <t \leq T),
\]
where ${\cal F}(t)$ is the $\sigma$-field generated by $\{(N_2(s), N(s)): s<t\}$.
We assume that
\begin{equation}\label{eq:aalen}
	\lambda(t)=\int_0^{t-} \mu(t/T,t-s) S(t/T,s) N_2(\d s),
\end{equation}
where $\mu(t/T,t-s)$ is the hazard function for the duration time in the hospital for an individual that entered at time $s$ and leaves at time $t$, and $S(t/T,s)$ is the survival function of duration-time-in-hospital for an individual that enters at $s$. That is, $S(t/T,s)$ is the probability that an individual who enters the hospital at time $s$, still remains at time  $t$, with $s < t $. This probability can be calculated in terms of $\mu(\cdot,\cdot)$ as
\begin{equation}
	S(t/T,s)=\exp\left\{-\int_s^{t}\mu(u/T,u-s)\d u\right\} = \prod_{s<u\leq t}\left\{1-\mu(u/T,u-s)\d u\right\}.
\end{equation}
{It is possible to view this problem within the framework of a Aalen's multiplicative intensity model as follows. We observe $n$ individuals in the interval $(0, T]$, and denote $N_{\cdot, i}(t)$ the process that counts 1 if individual $i$ has left the hospital (due to death or recovery) in $(0,t]$ and 0 otherwise. Let $\lambda_i(t)$ denote the corresponding intensity function ($i=1, \ldots, n$).\\ 
	\noindent Conditioned on the hospitalization history, which is the $\sigma$-field ${\cal F}(t)$, we observe the sequence of hospitalization times $\{\tau_i; i=1,\ldots, n\}$. Let $Y_i(t)$ be a random variable that takes the value 1 if individual $i$ has left the hospital before time $t$ and 0 otherwise. Then, the corresponding individual intensity rate is
	\[
	\lambda_i(t)=\mu(t/T,t-\tau_i)I(\tau_i<t)Y_i(t),
	\]
	which fits the Aalen's  multiplicative intensity model because $I(\tau_i<t)Y_i(t)$ is a predictable process (see \cite{Andersen:etal:93} and \cite{Nielsen:98}).\\
	When considering the aggregated process, we have that $N(t)=\sum_{i=1}^n N_{\cdot, i}(t)$, and if we replace $Y_i(t)$ by its expectation,  the integral formulation for the intensity function of the aggregated process is the expression given in \eqref{eq:aalen}.}

In Section \ref{sec:estim3y4} we will build an estimator of $\mu(t/T,t-s)$, for $0< s < t \leq T$. To do it we need observations of the process  $N(t,\d s)=\sum_{i=1}^nN_{\cdot, i}(t,\d s)$, where $N_{\cdot, i}(t,\d s)$ takes value 1 if the individual $i$, hospitalized in $(s, s+\d s]$, leaves at some time in the interval  $(s, t]$. Ideally we observe these processes but in our motivating application we only observe the marginal processes $N_2(t)$ and $N(t)$.

\section{Model estimation}\label{sec:estim}
\subsection{Estimating transition intensities from infection to infection} \label{sec:estim1}
We assume observations of the counting process $N_1(t)$ in the interval $(0, T]$. In our motivating application to the Covid-19 pandemic, this corresponds to the daily numbers of newly confirmed cases.\\
An iterative estimation scheme for the estimation of $\mu_1$ defined in Section \ref{sec:model1} is presented in the following. More details are in \cite{Gamiz:etal:25}.\\
\noindent At the $r$th iteration of the algorithm let $\widehat{\mu}_1^{(r-1)}$ denote an estimate of $\mu_1$ obtained at the preceding iteration.
For $r=1$ an initial guess $\widehat{\mu}_{1}^{(0)}$ of $\mu_1$ is used. The $r$th iteration of the algorithm consists of two steps: 
\begin{enumerate}
	\item  Construct a two-dimensional process $\widehat N_1^{(r)}(t,\d s)$ that approximates $N_1(t,\d s)$. This is done by using the estimator $\widehat{\mu}_1^{(r-1)}$ from  the previous iteration  and  by using the  observed process $N_1(t)$ {as follows}:
	\begin{equation}\label{eq:alpha1a}\widehat N_1^{(r)} (\d u, \d v)= \frac { \widehat \mu_1^{(r-1)} (u/T, u-v) }
		{\int_0^{u^-}\widehat \mu_1^{(r-1)} (u/T, u-w)  N_1(\d w)} N_1(\d v) N_1(\d u). \end{equation}
	\item Update $\widehat{\mu}_1^{(r-1)}$, by means of
	\begin{eqnarray}\label{eq:mu1}
		&&\hspace{-1cm} \widehat{\mu}_1^{{(r)}}(t/T,t-s)\\
		\nonumber &&= \frac{ \int_{0\leq v < u \leq T} {\rm D}_1(s,t,v,u) \ K_{1,b_1}\left((t-u)/{T}\right) K_{2,b_2}(t-s-(u-v)) \widehat N_1^{(r)}(\d u,\d v)}
		{ \int_{0\leq v < u \leq T}{\rm D}_1(s,t,v,u) \ K_{1,b_1}\left(({t-u})/{T}\right) K_{2,b_2}(t-s-(u-v)) N_1(\d v) \ \d u}.
	\end{eqnarray}
	where we have used the following notation:  $K_{j,b_j}(\cdot)=K_j(\cdot/b_j)/b_j$ , where $K_j$ ($j =1,2$) is a general one-dimensional kernel function, and, 
	\begin{equation*}
		{\rm D}_1(s,t,v,u)=1-\big ((t-u)/T,t-s-(u-v)\big )^T  {\rm {\bf A}}_1(t,s)^{-1}{\rm {\bf a}}_1(t,s) \quad (0 \leq s <t \leq T),
	\end{equation*} 
	with ${\rm {\bf a}}_1$  a vector function whose $l$th component is
	\begin{eqnarray*}\label{eq:a1}
		&& \hspace{-1cm}{\rm a}_{1,l}(t,s)=\int_{0\leq v < u \leq T}\! K_{1,b_1}\left(({t-u})/{T}\right) K_{2,b_2}(t-s-(u-v))
		\\ && \qquad   \qquad \times \left(({t-u})/{T}\right)^{\delta_{l1}}(t-s-(u-v))^{\delta_{l2}} N_{1}(\d v) \ \d u \qquad (l=1,2),
	\end{eqnarray*}
	being $\delta_{lh}=1$ if $l=h$, and 0, otherwise; and, with ${\rm {\bf A}}_1$  a matrix function with dimension $2 \times 2$ whose $(l,m)$-element is 
	\begin{eqnarray*}\label{eq:A1}
		\nonumber &&\hspace{-1.5cm} {\rm A}_{1,l,m}(t,s)=\int_{0\leq v < u \leq T}\! K_{1,b_1}\left(({t-u})/{T}\right)K_{2,b_2}(t-s-(u-v))
		\\ &&  \qquad \times \left(({t-u})/{T}\right)^{\delta_{l1}+\delta_{m1}}(t-s-(u-v))^{\delta_{l2}+\delta_{m2}} N_{1}(\d v) \ \d u \quad (l,m=1,2),
	\end{eqnarray*}
	with $\delta_{lh}$ defined above, and $\delta_{mh}=1$ if $m=h$ and 0 otherwise. 
\end{enumerate}
The estimator in \eqref{eq:mu1} is a ratio of a local linear estimator of occurrences and a local linear estimator of exposure, see \cite{Nielsen:98} that is a local linear generalisation of \cite{Nielsen:Linton:95}.
\subsection{Estimating transition intensities from infection to hospitalization}\label{sec:estim2}
Observations of the counting process $N_1(t)$ and $N_2(t)$, over the interval $[0 , T]$, with $N_2$ counting the number of hospital admissions are available. In the following, an iterative algorithm to estimate the transition rate from infected to hospitalized, $\mu_2$ is described (see \cite{Gamiz:etal:25} for further details). The procedure uses a similar approach as for the estimation of $\mu_1$, discussed in the last section. At the $r$th iteration of the algorithm let $\widehat{\mu}_2^{(r-1)}$ denote an estimation of $\mu_2$ obtained at the preceding iteration.
For $r=1$, an initial guess $\widehat{\mu}_{2}^{(0)}$ of $\mu_2$ is used. The $r$th iteration of the algorithm consists of two steps:  
\begin{enumerate}
	\item Construct a two-dimensional process $\widehat N_2^{(r)}(t,\d s)$ that approximates $N_2(t,\d s)$. This is done by using the estimator $\widehat{\mu}_2^{(r-1)}$ from  the previous iteration, and the  observed processes $N_1(t)$ and $N_2(t)$ as follows:
	\begin{equation*}\widehat N_2^{(r)} (\d u, \d v)= \frac { \widehat \mu_2^{(r-1)} (u/T, u-v) }
		{\int_0^{u^-}\widehat \mu_2^{(r-1)} (u/T, u-w)  N_1(\d w)} N_1(\d v) N_2(\d u). 
	\end{equation*}
	\item  Update $\widehat{\mu}_2^{(r-1)}$, by means of
	\begin{eqnarray}\label{eq:mu2}
		&&\hspace{-2cm} \widehat{\mu}_2^{{(r)}}(t/T,t-s)\\
		\nonumber	&&= \frac{ \int_{0\leq v < u \leq T} {\rm D}_1(s,t,v,u) \ K_{1,b_1}\left(\frac{t-u}{T}\right) K_{2,b_2}(t-s-(u-v)) \widehat N_2^{(r)}(\d u,\d v)}
		{ \int_{0\leq v <u \leq T}{\rm D}_1(s,t,v,u) \ K_{1,b_1}\left(\frac{t-u}{T}\right) K_{2,b_2}(t-s-(u-v)) N_1(\d v) \ \d u}.
	\end{eqnarray}
\end{enumerate}
The performance of the estimator in \eqref{eq:mu2} can be discussed by using similar arguments as that of $\widehat \mu_1$ in the last section {(compare with equation \eqref{eq:mu1})}. Here ${\rm D}_1$ is defined as in Section \ref{sec:estim1}.
\subsection{Estimating the transition rate from hospitalization to death or recovery}\label{sec:estim3y4}
Let us assume that, in addition to $N_2(t)$, we observe the marginal processes $N_3(t)$ and $N_{4}(t)$, which represent daily hospital discharges and hospital deaths, respectively. Let $\mu_{3}(t,t-s)$ denote the recovery hazard function at time $t$ for a subject that entered the hospital at time $s$, and $\mu_{4}(t,t-s)$ the mortality hazard function at time $t$ for a subject that entered the hospital at time $s$. In this section, an iterative estimation scheme for the estimation of $\mu_3$, and $\mu_4$ is introduced. \\
Let us denote $N(t)=N_3(t)+N_4(t)$ the total number of patients leaving the hospital at time $t$, and $\mu=\mu_3+\mu_4$, the hazard function for duration time in hospital, regardless the patient leaves the hospital due to death or clinical discharge. Let $Y(t,s)$ the total number of individuals that enter the hospital at time $s$ and still remain hospitalized on the day $t$ ($0\leq s<t \leq T$). Observations of this two-dimensional process are not available on the dataset.\\
At the $r$th iteration of the algorithm let $\widehat{\mu}^{(r-1)}$ denote an estimation of $\mu$ obtained at the preceding iteration. For $r=1$ an initial guess $\widehat{\mu}^{(0)}$ is used as the starting estimate of $\mu$. The $r$th iteration of the algorithm consists of two steps: 
\begin{enumerate}
	\item Construct a two-dimensional process $\hat N^{(r)}(t,\d s)$ that approximates $N(t,\d s)$ and a two-dimensional process $\hat Y^{(r)}(t,s)$ that approximates $Y(t,s)$. This is done by using the estimator $\widehat{\mu}^{(r-1)}$ from  the previous iteration  and  by using the  observed processes $N_2(t)$ and $N(t)$. Let us denote 
	\[
	\widehat{S}^{(r-1)}(t/T,s)=\prod \limits_{s < u \leq t}\! \left\{1-\widehat{\mu}^{(r-1)} (u/T, u-s)\d u\right\}.
	\]
	the estimated probability that a subject hospitalized at time $s$ still remains in hospital at time $t$. We define
	\begin{equation}\label{eq:N}
		\hat N^{(r)} (\d u, \d v)= \frac { \widehat{S}^{(r-1)}(u/T,v) \widehat{\mu}^{(r-1)} (u/T, u-v) N_2(\d v)}
		{\displaystyle {\int_0^{u-}}\! \widehat{S}^{(r-1)}(u/T,w) \widehat{\mu}^{(r-1)} (u/T, u-w)  N_2(\d w)}   N(\d u).
	\end{equation}
	and
	\begin{equation}\label{eq:Y}
		\hat Y^{(r)} (u,v) \d v \d u= \frac {\widehat{S}^{(r-1)}(u/T,v)  N_2(\d v)} 
		{{\displaystyle {\int_0^{u-}}}\!\widehat{S}^{(r-1)}(u/T,w) N_2(\d w)} Y(u) \d u,
	\end{equation}
	with $Y(u)=N_2(u)-N(u) $, is the number of individuals remaining in hospital at time $u$.
	\item The estimator of   $ N (\d u, \d v)$ given in \eqref{eq:N} and the estimator of $Y(u,v)$ given in \eqref{eq:Y} are used to update $\widehat{\mu}^{(r-1)}$ as follows. Define
	\begin{eqnarray}\label{eq:mu_r}
		&&\hspace{-1cm} \widehat{\mu}^{(r)}(t/T,t-s)=\\
		\nonumber	 && \frac{ \int_{0\leq v < u \leq T} {\rm D}_2(s,t,v,u) \ K_{1,b_1}\left((t-u)/T\right) K_{2,b_2}(t-s-(u-v)) \hat N^{(r)}(\d u,\d v)}
		{ \int_{0\leq v < u \leq T}{\rm D}_2(s,t,v,u) \ K_{1,b_1}\left((t-u)/T\right) K_{2,b_2}(t-s-(u-v)) \hat Y^{(r)}(u,v) \d v  \d u},
	\end{eqnarray}
	where we have used the following notation:  
	\begin{equation*}
		{\rm D}_2(s,t,v,u)=1-\big ((t-u)/T,t-s-(u-v)\big )^T  {\rm {\bf A}}_2(t,s)^{-1}{\rm {\bf a}}_2(t,s) \quad (0 \leq s <t \leq T).
	\end{equation*} 
	Here ${\rm {\bf a}}$ be a vector function whose $l$th component is
	\begin{eqnarray*}
		&&\hspace{-2cm} {\rm a}_{2,l}(t,s)=\int_{0 \leq v < u \leq T} K_{1,b_1}\left((t-u)/T\right) K_{2,b_2}(t-s-(u-v))
		\\ && \times \left((t-u)/T\right)^{\delta_{l1}}(t-s-(u-v))^{\delta_{l2}} Y(u,v) \d v  \d u \quad (l=1,2),
	\end{eqnarray*}
	being $\delta_{lh}=1$ if $l=h$, and 0, otherwise. Also, ${\rm {\bf A}}_2$ is a matrix function with dimension $2 \times 2$ whose $(l,m)$-element is given by
	\begin{eqnarray*}\label{eq:A2}
		&& \hspace{-1.5cm} {\rm A}_{2,l,m}(t,s)=\int_{0 \leq v < u \leq T}  K_{1,b_1}\left((t-u)/T\right)K_{2,b_2}(t-s-(u-v))
		\\ && \qquad \times  \left((t-u)/T\right)^{\delta_{l1}+\delta_{m1}}(t-s-(u-v))^{\delta_{l2}+\delta_{m2}} Y(u,v) \d v  \d u \quad (l,m=1,2),
	\end{eqnarray*}
	with $\delta_{lh}$ defined above, and $\delta_{mh}=1$ if $m=h$ and 0 otherwise. \\
	As denoted previously, for $j =1,2$, $K_{j,b_j}(\cdot)=K_j(\cdot/b_j)/b_j$, where $K_j$ is a general one-dimensional kernel function.
\end{enumerate}
Finally, two types of outcome are considered. Specifically, we estimate the transition rate from-hospitalized-to-recovery, $\mu_{3}(t/T,w)$, and the transition rate from-hospitalized-to-death, $\mu_{4}(t/T,w)$ , respectively, as follows. Let $\widehat{\mu}$ be the solution of the above iterative procedure based on  $\hat N$ and $\hat Y$ obtained at the last iteration. Let us define the following equation, similar to \eqref{eq:N},
\begin{equation}\label{eq:N_j}
	\hat N_{j} (\d u, \d v)= \frac { \widehat{S}(u/T,u-v)\widehat{\mu} (u/T, u-v) N_2(\d v)  }
	{\displaystyle {\int_0^{u-}}\! \widehat{S}(u/T,u-v) \widehat{\mu} (u/T, u-w)  N_2(\d w)}  N_{j}(\d u) \quad (j=3,4).
\end{equation}
Then
\begin{eqnarray}\label{eq:mu_j}
	&&\widehat{\mu}_{j}(t/T,t-s) = \\
	\nonumber&&\frac{ \int_{0\leq v < u \leq T} {\rm D}_2(s,t,v,u) \ K_{1,b_1}\left((t-u)/T\right) K_{2,b_2}(t-s-(u-v)) \hat N_{j}(\d u,\d v)}
	{ \int_{0\leq v < u \leq T}{\rm D}_2(s,t,v,u) \ K_{1,b_1}\left((t-u)/T\right) K_{2,b_2}(t-s-(u-v)) \hat Y(u,v) \d v  \d u} \quad (j=3,4).
\end{eqnarray}
\medskip

\begin{remark}\
	
	\begin{enumerate}
		\item [(i)] 	In the Appendix \ref{app:theo1} we will argue that the just described iterative estimator where the iteration is stopped if the difference between two consecutive  solutions is below a threshold of order $o(1)$, is a consistent estimator. 
		
		\item[(ii)] When choosing the bandwidths $b_1$ and $b_2$ for the local linear marker-dependent hazard estimator were selected by the cross-validation method of \cite{Gamiz:etal:13}, which performed well in the applications considered here. We also examined a refined version of cross-validation proposed by \cite{Mammen:etal:11}, adapted to one-dimensional local linear hazards by \cite{Gamiz:etal:16} and to marker-dependent hazards by \cite{Gamiz:etal:13}. Although this method is theoretically more robust, in our setting cross-validation and Do-validation yielded similar results, and we therefore adopted the simpler cross-validation approach.
		\item[(iii)]The performance of the proposed missing-link marker-dependent hazard estimator was assessed by finite-sample simulations. The estimator performed well, with accuracy improving as the measurement error decreased when larger sample sizes were considered. As expected, the performance was inferior to that obtainable under full information without the missing-link survival problem. Details of the simulation study are given in Appendix~\ref{app:simulations}. Under full information we observe independent and identically distributed pairs of processes $(N_{2,1}(\d s),N_{\cdot, 1}(t,\d s)), \ldots,(N_{2,n}(\d s),N_{\cdot, n}(t,\d s))$  ($s<t \leq T$).
	\end{enumerate}
\end{remark}

\section{Forecasting}\label{sec:forecast}
\subsection{Forecasting the number of infections using expert knowledge.}\label{sec:C1}
The main view of forecasting taken in this paper is related to the considerations of Section \ref{sec:full_system}. 
When looking at the full system described in Section \ref{sec:full_system} for the Covid-19 case of France, one can take that point of view that all transitions except that one from the infection indicator to the infection indicator can be described via slowly moving continuous development over time. In other words: except for this one transition, it makes sense, at any given date, to forecast the immediate future structure of these transitions from the immediate past. \\
With respect to the infection process, \cite{Gamiz:etal:25} introduced a quantity $C_{1,h}$ indicating whether the future, that is $(T,T+h]$ for $h>0$, will be different or equal to the immediate past. Specifically, if $C_{1,h}=1$ then can forecast in a future time interval $(T, T+h]$ based on the immediate past. \\
The key feature of our forecasting approach is that monitoring and prediction are feasible when $C_{1,h}=1$. In this case, a dynamic statistical methodology can be applied using a surprisingly simple data collection reflecting what can be considered ``available data'' in many countries. However, pandemics are characterized by multiple change-points in severity.  Therefore, effective monitoring and forecasting require a dynamic perspective combined with knowledge of $C_{1,h}$. This task needs expert input specific to countries or local regions.  Such expertise cannot be dismissed via a statistical analysis based on simplified available data. The role of experts is to focus on forecasting and understanding $C_{1,h}$ rather than being responsible for a full statistical model.  \\
A suitable value of $C_{1,h}$ at time $T$ informed by expert knowledge, enables our methodology to forecast the number of new infected in $(T, T+h]$ for a fixed $h>0$. From those forecast, we can predict the number of hospitalized to thus forecast number of deaths in hospitals as well as the number of recoveries in the interval $(T, T+h]$.

\subsection{Forecasting the total number of deaths.}\label{sec:C2}
In this paper we introduce an additional indicator $C_{2,h}$ to predict the total number of deaths, both inside and outside hospital, i.e. $N_{\rm out}+ N_{\rm in}$ as defined above, based on forecasted number of infected provided by  $C_{1,h}$ and the dynamic model illustrated in Figure \ref{fig:diagram}.  \\
For any $t$ in the observation interval $(0, T]$, we consider the estimator of the ratio $F(t)$ described in Section \ref{sec:deaths}, denoted by $\widehat{F}(t)$, for all $t \in (0,T]$.\\
\noindent Let us fix the forecast horizon at time $T+h$, for an $h>0$. To extrapolate the ratio $F(T+s)$, for $0< s \leq h$, we assume that the ratio $F$ at the end of the forecasting period (i.e. $T+h$) is the more recent estimate multiplied by  $C_{2,h}$, and it varies linearly in between, that is,
\begin{equation}\label{eq:fore}
	\widetilde F(T+s)=\widehat F(T) \times \left [1+ (C_{2,h}-1)\frac{s}{h}\right ] \quad (0< s \leq h).
\end{equation}

Under the assumption that the ratio of densities is constant and equal to the most recent estimation, that is, $\widetilde{F}(T+s)=\widehat{F}(T)$ ($0<s \leq h$), it has to be chosen $C_{2,h}=1$. On the other hand, we can consider an infeasible in practice procedure to choose the optimal value of the constant by optimizing the predicted number of daily deaths in the target forecasting interval $(T, T+h]$. \\
For simplicity in the following we remove the subscript $h$ from the notation and write the above indicators as $C_1$ and $C_2$.

\subsection{Uncertainty  quantifications of predictions}\label{sec:uncertainty}
We propose a novel bootstrap method to quantify uncertainty by constructing prediction bands for fixed values $C_{1}$ and $C_{2}$. The method assumes  a more general model than the Hawkes model for the infection process, which is able to account for the overdispersion observed in the dataset; see \cite{Gamiz:etal:25}. For transitions from infection to hospitalization, uncertainty is assumed to be adequately captured by model \eqref{eq:mu2}, while for transitions outside the hospital, event occurrences follow the specification in \eqref{eq:mu_j}. No substantial discrepancies between the data and these model assumptions were observed.

In the following, we outline the proposed method. A detailed description of the algorithm is given in Appendix \ref{app:bootstrap}. \

First, we calculate the empirical index of dispersion based on observations of the infection process as follows. Assume we observe $N_{1,0}, N_{1,1}, \ldots, N_{1,T}$, where $N_{1,i}$ is the number of infections at the $i$th day ($i=0,2,\ldots, T$). We can assess potential overdispersion in the data by evaluating the dispersion index given by
\begin{equation}\label{eq:gamma}
	\gamma=\frac{1}{T}\sum_{i=1}^T\frac{(N_{1,i}-\lambda_{1,i})^2}{\lambda_{1,i}},
\end{equation}
which compares the empirical variance to the expected variance under the Hawkes model assumption. Here $\lambda_{1,i}$ is the expected number of infections at the $i$th day, given $N_{1,0},\ldots, N_{1,i-1}$, that is $\lambda_{1,i}=\sum_{j=0}^{i-1}N_{1,j}\mu_1(i/T,i-j)$.

For the covid-19 dataset analysed in Section \ref{sec:application}, the factor in \eqref{eq:gamma} is estimated to be $\gamma >1$, indicating substantial overdispersion relative to the Hawkes model. To account for the overdispersion present in the infection process, we quantified the uncertainty of our predictions using a bootstrap method, in which bootstrap samples were generated as follows.
\begin{enumerate}
	\item First, we simulate new infections using a bootstrap procedure that accounts for the possibility of super-spreader events. The procedure begins by selecting a parameter $k > \gamma$, where $\gamma$ is the variance factor estimated in \eqref{eq:gamma}. We define  $\beta = (\gamma - 1)/(k(k-1))$, and $\alpha = 1 - k \beta$.  
	
	The expected number of infected on the day $i$ is $\lambda^*_{1,i} = N^*_{1,0} \, \hat{\mu}_1(i/T,i) + N^*_{1,1} \, \hat{\mu}_1(i/T,i-1) + \dots + N^*_{1,i-1} \, \hat{\mu}_1(i/T,1)$. Then we simulate  
	$N^*_{1,\alpha,i} \sim \text{Pois}(\alpha \lambda^*_{1,i}),$ and $N^*_{1,\beta,i} \sim \text{Pois}(\beta \lambda^*_{1,i})$, and the total number of infections for day $i$ is given by  $N^*_{1,i} = N^*_{1,\alpha,i} + k \, N^*_{1,\beta,i}$.  
	
	The parameters $\alpha$ and $\beta$ are chosen so that the conditional mean of the simulated infections matches the expected value, while the conditional variance matches $\gamma$ times the expected value. In our implementation, $k$ is set to $\gamma + 1$, with the estimated variance factor $\gamma = 65.3$ for daily infection data in France.  
	
	
	\item The bootstrap procedure is extended to generate hospitalizations. After simulating the infection counts, the number of hospitalizations on each day is generated as a Poisson random variable, with its mean computed as a weighted sum of past  simulated infections. Specifically, given $N^*_{1,0}, N^*_{1,1}, \ldots, N^*_{1,i}$, generate $N^*_{2,i} \sim Pois(\lambda^*_{2,i})$, with $\lambda^*_{2,i}=N^*_{1,1} \hat{\mu}_{2}(i/T,i)+ N^*_{1,2} \hat{\mu}_2(i/T,i-1)+\ldots +N^*_{1,i} \hat{\mu}_2(i/T,1)$. This ensures that hospitalizations are appropriately linked to prior infections, maintaining consistency with the observed infection-hospitalization relationship.

	
	\item After simulating hospitalizations, the bootstrap procedure continues by generating daily numbers of recoveries and hospital deaths. This step accounts for patients who remain hospitalized on each day.  
	Given $N^*_{2,0}, N^*_{2,1},\ldots, N^*_{2,i}$, we define the expected number of patients who recover on day $i$ after having spent a total of $d$ days in hospital as  $\lambda^*_{3,i}(d) = N^*_{2,i-d+1} \, S(i/T, i-d+1) \, \hat{\mu}_3(i/T, d)$, where $S(i/T, \cdot)$ denotes the survival function of hospital duration for an individual who remains hospitalized at calendar time $i$, regardless of the admission time. This function accounts for both deaths and clinical discharges, and is computed as the product of the probabilities of not recovering or dying on the  days previous to $i$, that is,
	\[
	S(i/T,i-d+1) = \prod_{j=1}^{d-1} \left[1 - \hat{\mu}((i-j)/T,j)\right], (d=1,\ldots, i),
	\]
	with $\hat{\mu} = \hat{\mu}_3 + \hat{\mu}_4$. It is assumed that $S(i/T,i) = 1$ ($i=1,\ldots, T$). \\
	Similarly, the expected number of patients who die in hospital after a stay of $d$ days is given by  $\lambda^*_{4,i}(d) = N^*_{2,i-d+1} \, S(i/T, i-d+1) \, \hat{\mu}_4(i/T, d)$.
	On day $i$, the number of recoveries after a hospital stay of $d$ days is simulated as $N^*_{3,i,d} \sim Pois(\lambda^*_{3,i}(d))$, and he number of deaths after a hospital stay of $d$ days is simulated as $N^*_{4,i,d}\sim Pois(\lambda^*_{4,i}(d))$, respectively.
	Finally, regardless of the duration of stay in hospital, the total number of recoveries on day $i$ is $N^*_{3,i} = \sum_{d=1}^{i} N^*_{3,i,d}$, while the total number of deaths on day $i$ is $N^*_{4,i} = \sum_{d=1}^{i} N^*_{4,i,d}$.
	
	\item 
	For each bootstrap sample consisting of $\{(N^*_{1,i},N^*_{2,i},N^*_{3,i},N^*_{4,i}), i=0,1,\ldots,T\}$, we obtain forecasts of the number of infections, the number of hospitalized patients, and the total number of deaths (both inside and outside the hospital), using fixed values of $C_1$  and $C_2$ and following the specifications of Sections \ref{sec:C1} and \ref{sec:C2}. 
	We apply the forecasting procedure described in Section \ref{sec:C1} to obtain the predicted numbers of infections and hospitalizations over the forecasting interval. By repeating this process many times, we generate a large set of possible trajectories for infections and hospitalizations, from which prediction intervals are computed using the 2.5\% and 97.5\% quantiles. A complete data analysis of these two indicators is provided in \cite{Gamiz:etal:25}.\\
	Applying the forecasting procedure described in Section \ref{sec:C2} to each bootstrap sample of hospital deaths generated by the above method, we obtain the predicted total number of deaths in the forecasting interval. For this, we incorporate the observed number of deaths occurring outside the hospital. By repeating the process many times, we generate a large set of possible trajectories for the total number of deaths, from which prediction intervals are computed using the 2.5\% and 97.5\% quantiles of the predicted numbers of infections and hospitalizations.
\end{enumerate}


\section{Estimating with available data. The Covid-19 case of France}\label{sec:application} 
\subsection{Preliminary}

In this section we present our main ideas on monitoring a developing pandemic based on available data. We use the recent Covid-19 pandemic and the country France as our case study. We consider official data between 18th March 2020 and 3rd January 2022. We walk through our modelling principles and the new mathematical statistical inventions necessary to implement our new approach.  

The raw data exhibit pronounced day-of-week effects, with fewer reported cases on weekends than weekdays. While some of this variation may reflect changes in transmission related to human behaviour, it is more likely attributable to reporting delays, slower case confirmations, and delayed hospital discharges over weekends. To mitigate the impact of these fluctuations on our predictions, we preprocess the data following the method of \cite{Koyama:et:al:21} prior to analysing the daily case sequence; further details are provided in Appendix~\ref{app:koyama}.

\subsection{Estimation of the rate of transition from hospitalized to recovered or dead }\label{sec:hospital}

Time spent in hospital is a dynamic concept in the sense that it depends on the particular date an individual is admitted to hospital. In our model $\mu(t/T,t-s)$, we mean by $s <t$ two different calendar times. That is, $s$ refers to the time of admission to hospital whereas $t$ is the time a patient leaves the hospital, and so, $t-s$ is the duration of hospitalization. 

\begin{figure}[ht]
	\centering
	\includegraphics[width=1\textwidth]{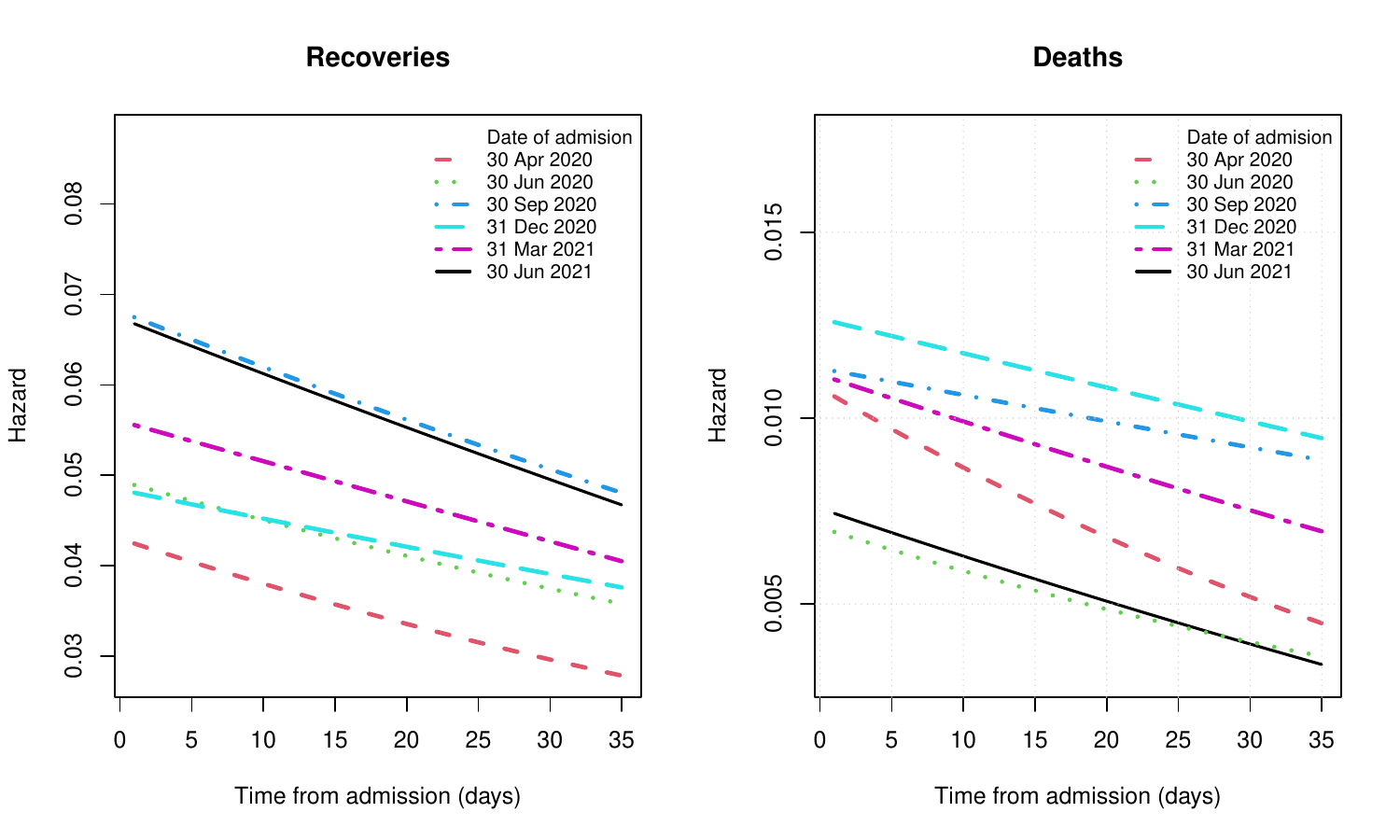}
	\caption{Hazard rate of time spent in hospital for individuals entering the hospital at different dates in the period from April 2020 to June 2021. Left panel: Hazard rate for time since admission until hospital discharge. Right panel: Hazard rate of time since admission until death.} \label{fig:lifetime-in-hospital}
\end{figure}

\begin{figure}[ht]
	\centering
	\includegraphics[width=1\textwidth]{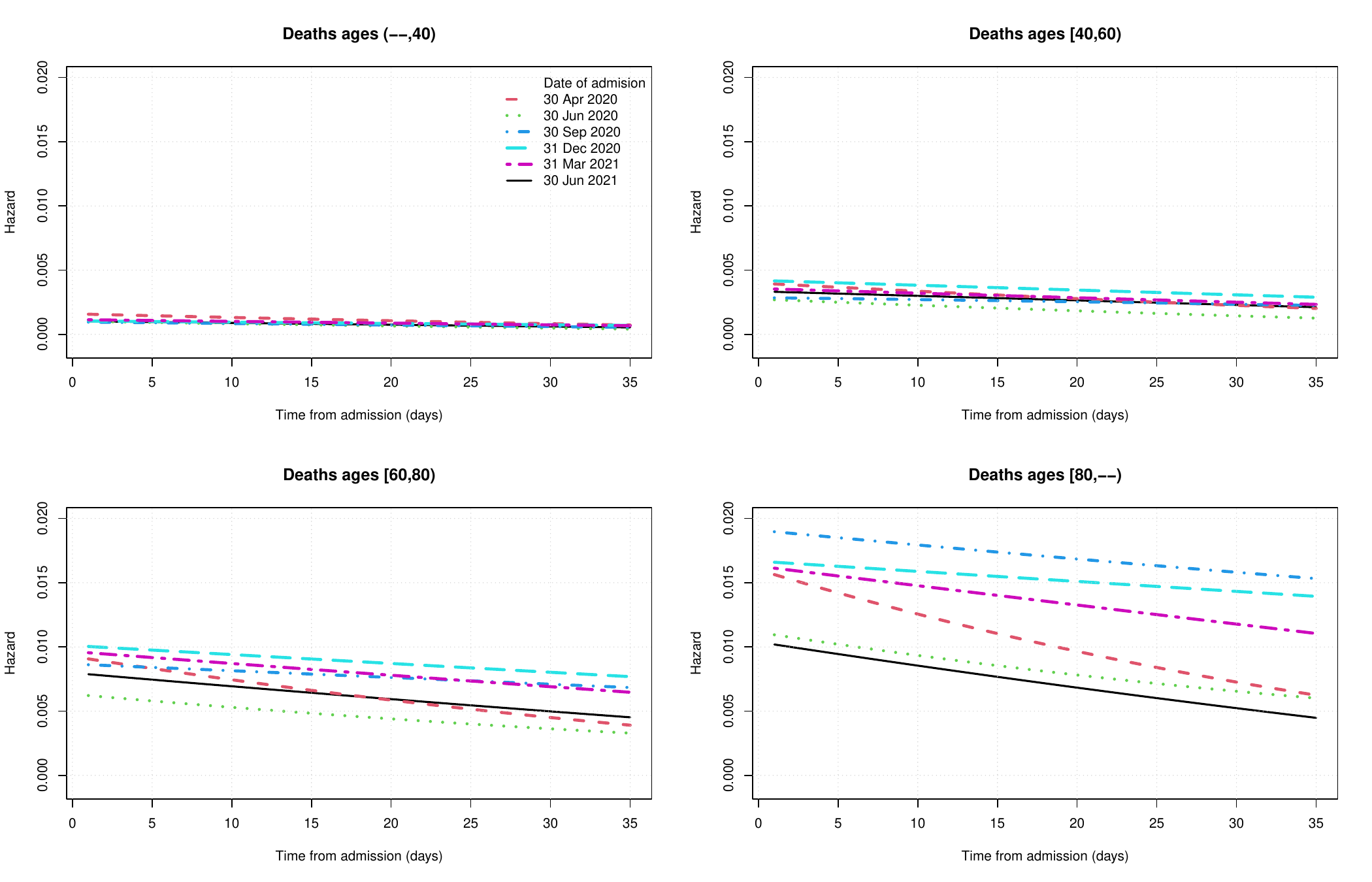}
	\caption{Hazard rate of time since admission until death for individuals entering the hospital at different dates in the period from April 2020 to November 2021. } \label{fig:hazard-ages}
\end{figure}

\begin{figure}[ht]
	\centering
	\includegraphics[width=1\textwidth]{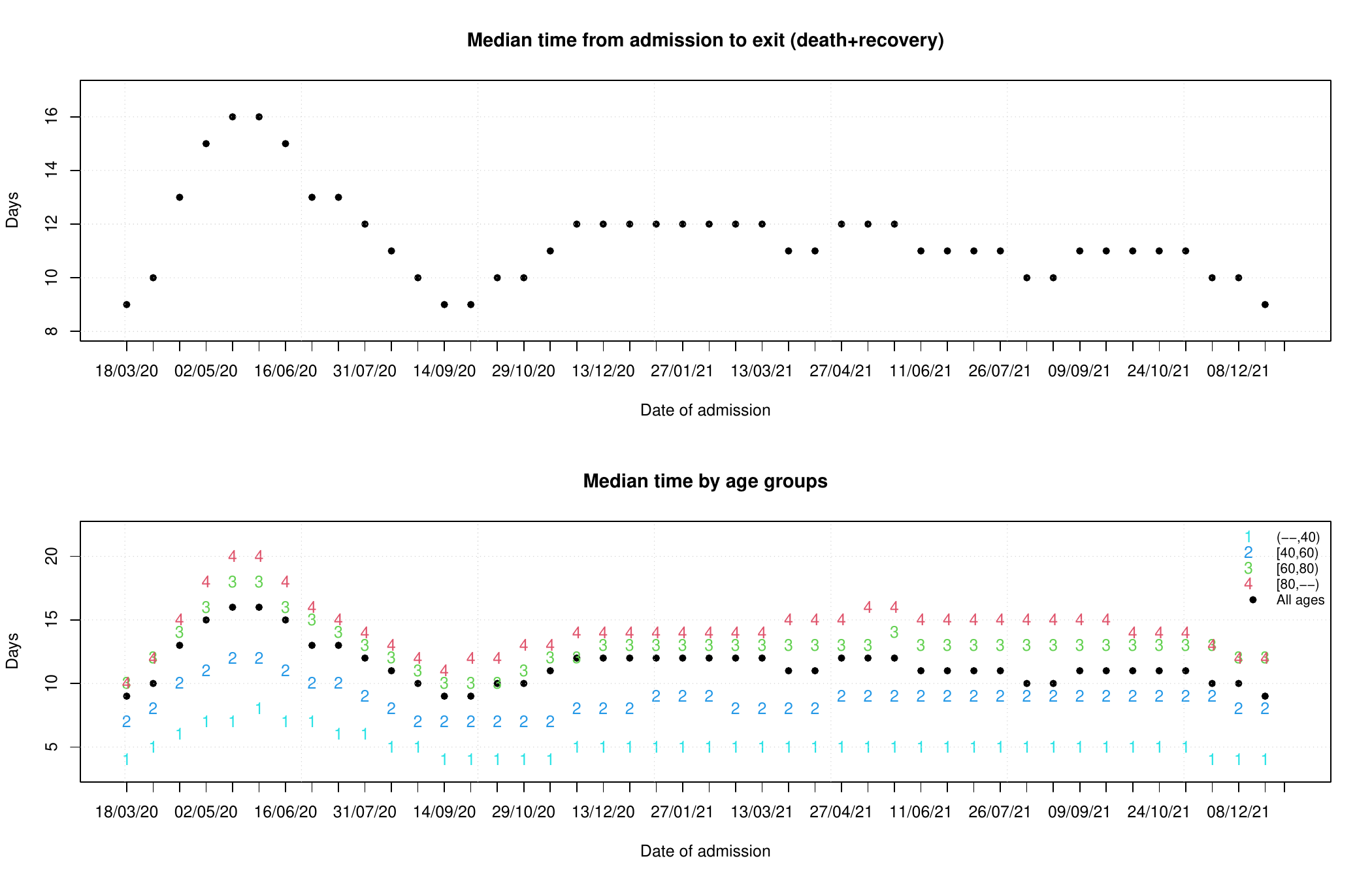}
	\caption{Median of the time spent in hospital by date of admission from March 2020 to January 2022. } \label{fig:mediantime}
\end{figure}

\begin{figure}[ht]
	\centering
	\includegraphics[width=1\textwidth]{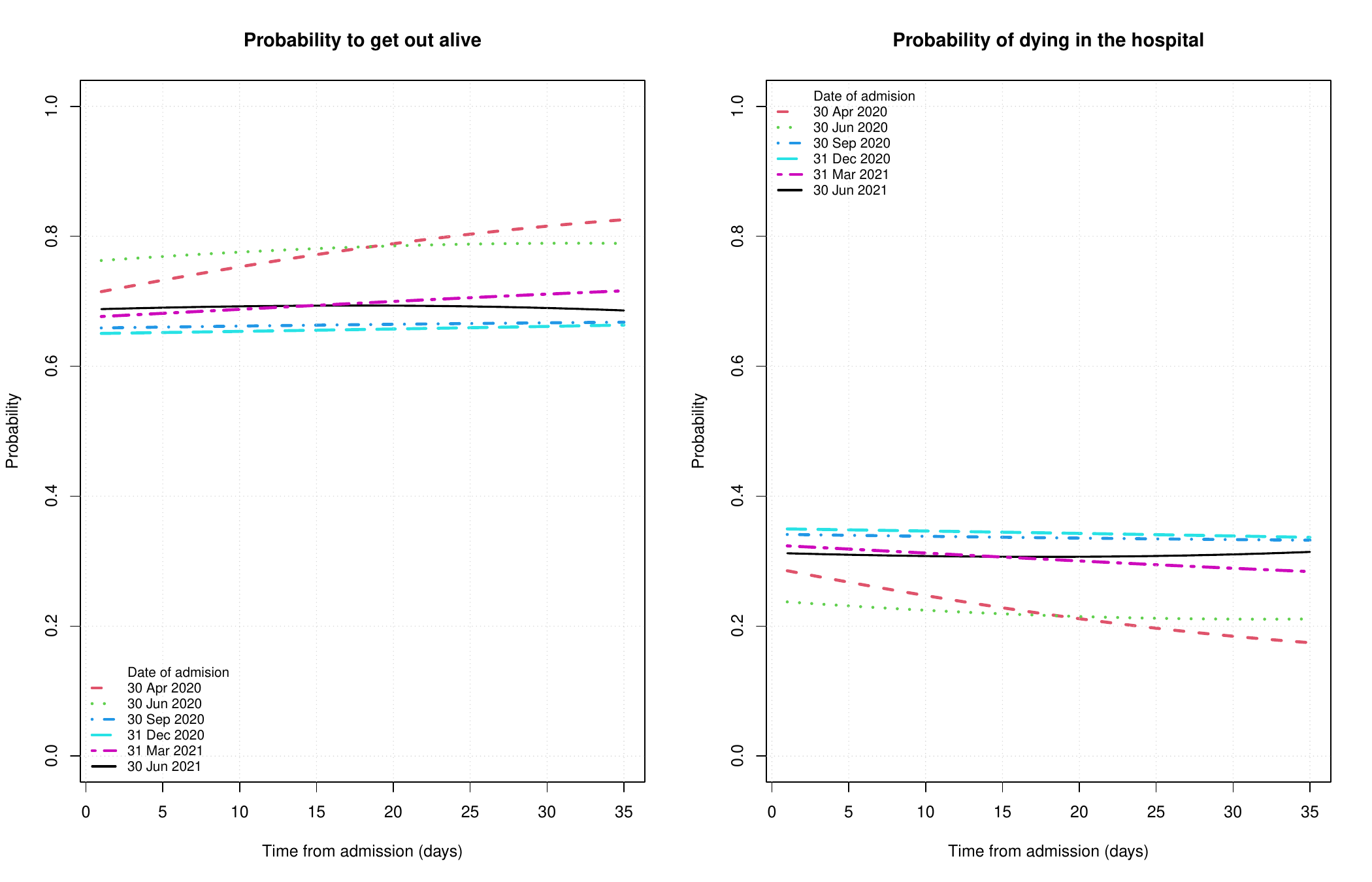}
	\caption{Probability of outcome by cause specific from March 2020 to January 2022. Left panel: Probability of leaving the hospital due to recovery. Right panel: Probability of leaving the hospital due to death.} \label{fig:probs}
\end{figure}

\begin{figure}[ht]
	\centering
	\includegraphics[width=1\textwidth]{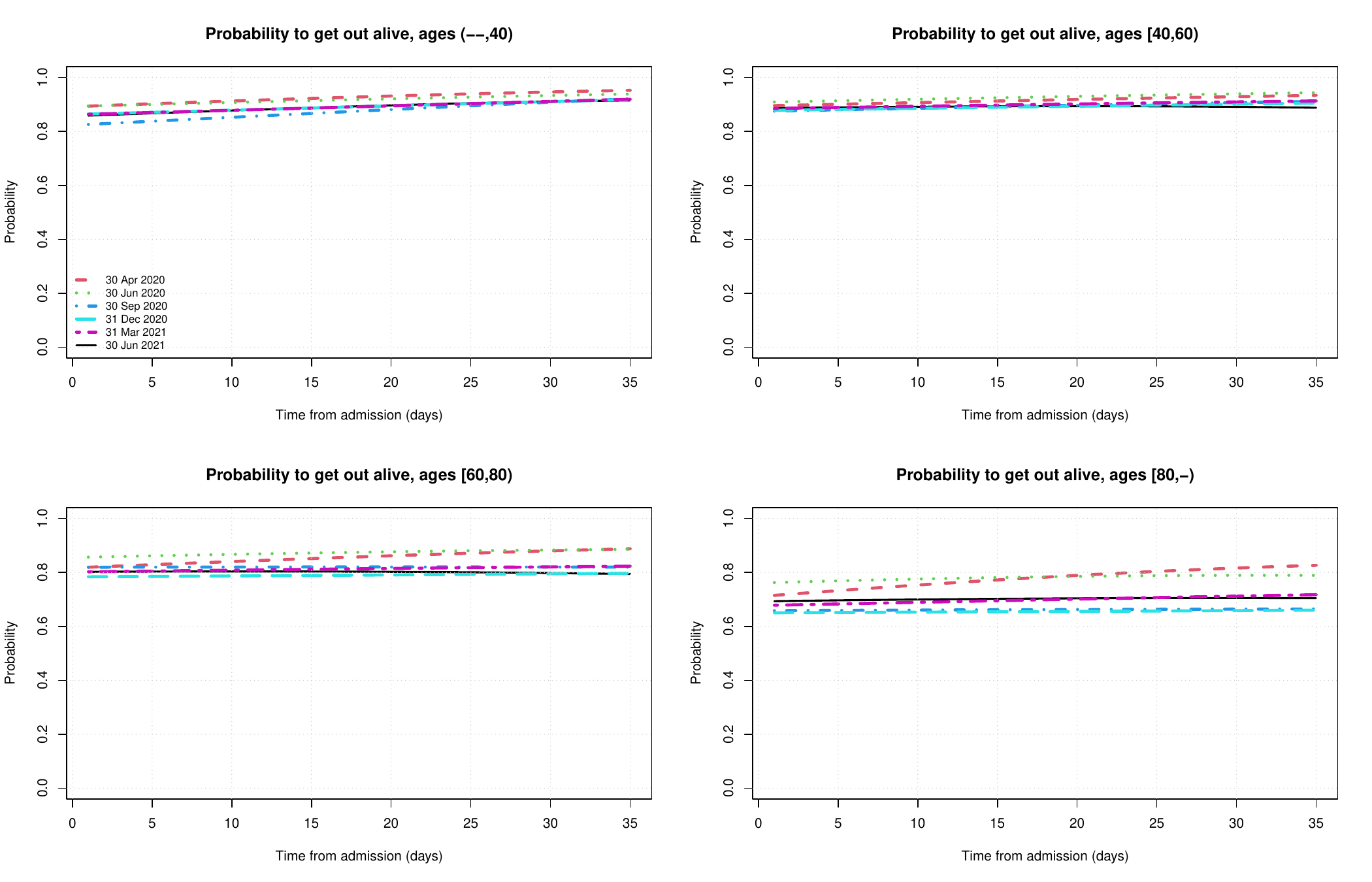}
	\caption{Probability of leaving the hospital due to recovery by age groups.} \label{fig:alive_ages}
\end{figure}

\begin{figure}[ht]
	\centering
	\includegraphics[width=1\textwidth]{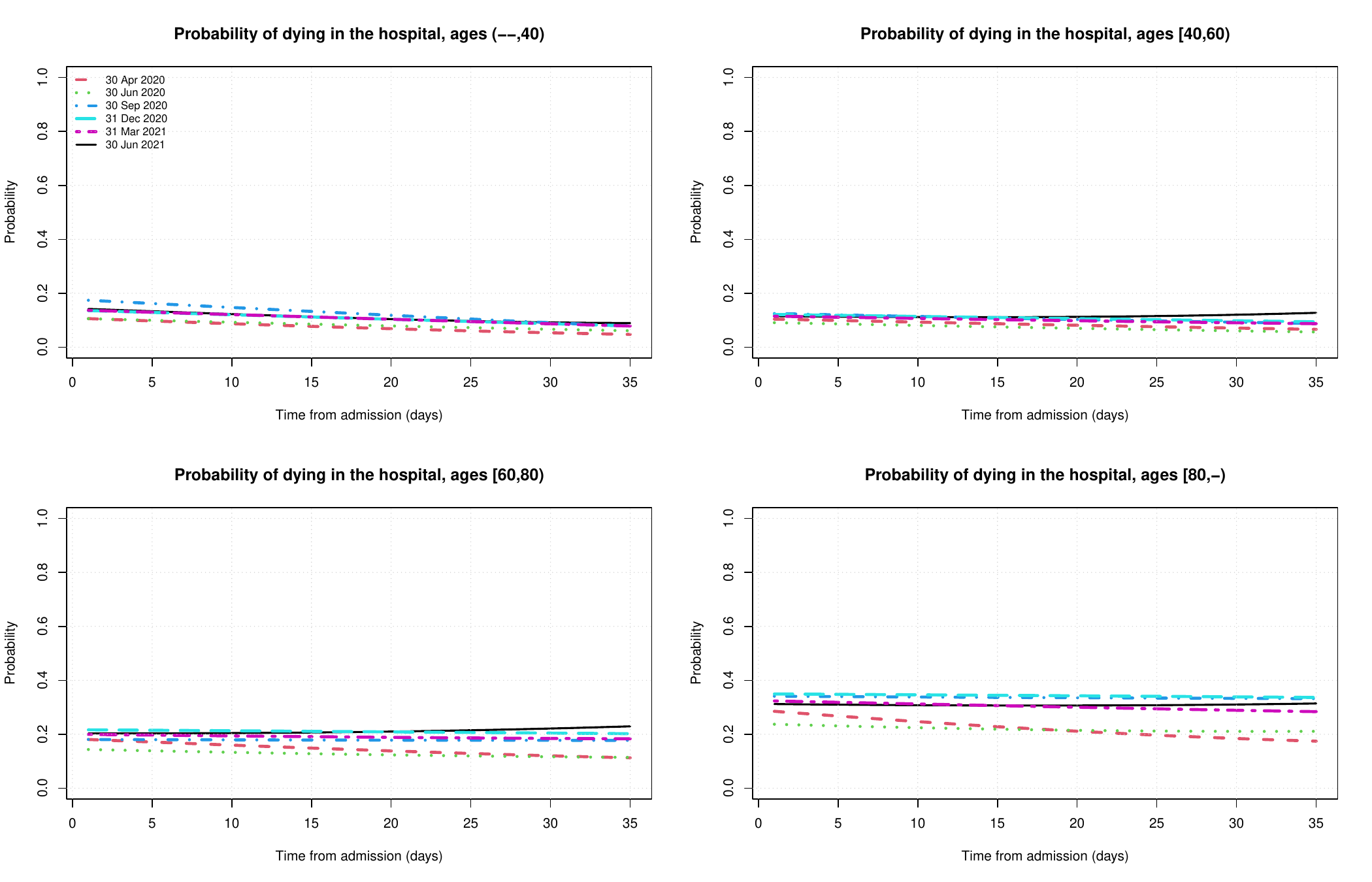}
	\caption{Probability of leaving the hospital due to death by age groups.} \label{fig:death_ages}
\end{figure}

Figures \ref{fig:lifetime-in-hospital} $-$ \ref{fig:death_ages} are designed to visualize the dynamics transitions outside hospital, during the French Covid-19 pandemic. To aid in understanding the situation, we have chosen to represent these transition rates from a perspective different from that provided by the functions $\mu_3(t/T,\cdot)$ and $\mu_4(t/T,\cdot)$ recoveries and deaths, respectively, which is the notation that we mainly use in the paper. 
Specifically, for $s$ fixed, we denote by $\alpha_3(s,t)$ the recovery rate at time $t$ of a process that initiated at time $s$. Thus, we have $\alpha_3(s,t) = \mu_3(t/T, t-s) \ $ ($0 < s < t \leq T$). Similarly, we denote  by $\alpha_4(s,t)$ the mortality rate at time $t$ for an individual infected at time $s$, and then we have $\alpha_4(s, t) = \mu_4(t/T, t-s) \ $ ($0 < s < t \leq T$).
With this notation,
Figure \ref{fig:lifetime-in-hospital} displays the estimations of the transition function from being in hospital to recovery, $\alpha_{3}$ (left panel), or death, $\alpha_{4}$ (right panel), for individuals entering at different dates, that is $s$. \\  Specifically, the red dashed curve in Figure \ref{fig:lifetime-in-hospital} represents the rate at which individuals admitted to hospital on 30st of April, 2020 recovered as a function of duration time in hospital. The plot shows curves corresponding to several values of the first dimension. For instance the solid black line in the left panel represents the hazard function of the duration until recovery for patients who enter the hospital on 30th June 2021. The hazard function can be interpreted as  the probability a patient leaves the hospital due to recovery conditioned on the duration of his/her stay. The solid black line shows a decreasing tendency as time in hospital for these patients passes. 

In concrete, we see that about 6.8\% of patients that entered the hospital on the 30th June 2021 received clinical discharge on the same day. 
Also, we can say that after a stay of 10 days, a person who was admitted to hospital on the 30th June 2021 has a probability about 0.06 of recovery, while the probability of recovery was barely of 0.038 on the 10th May 2020 for patients admitted on the 30th April 2020 (see dashed red line). 

The graph on the right panel of Figure \ref{fig:lifetime-in-hospital} shows that for people  just admitted in hospital on the 30th of April 2020 the risk of death is about 1.1\% and decreases with the time in hospital. The risk is about 0.75\% for people arriving to the hospital two months later.  
This can be an indicator of the collapse of the hospitals in the months immediately after the pandemic outbreak (March and April).

The probabilities of hospitalization, recovery, or death given infection may vary over calendar time. Factors such as political decisions, hospital resource availability, staff fatigue, and the evolving characteristics of the pandemic all contribute to this temporal variation. Consequently, all definitions in this paper are time-dependent, and our mathematical framework accommodates such dynamics. The calendar-time dependency of our two-dimensional marker-dependent hazard is a key component for fully flexible, dynamic modelling of a developing pandemic.
In the context of Covid-19 in France, improvements in clinical experience and understanding of the disease from the first wave in spring 2020 to the second wave in autumn 2020 are evident. As illustrated in the left panel of Figure~\ref{fig:lifetime-in-hospital}, the conditional probability of hospital discharge, given a patient has been hospitalized for $x$ days, increased over this period for all $x$. For instance, for patients admitted the previous day ($x=1$) the probability of leaving hospital due to recovery ranges from 4.2\% on the 30th April to 6.8\% on the 30th September 2020.

The average daily number of new hospitalizations was approximately four times higher at the end of October 2020 than in April 2020. This variation implied  a substantial  pressure on hospitals in October affecting the overall chances of recovering from the infection while in hospital. As shown in  Figure \ref{fig:lifetime-in-hospital}, the probability of discharge alive clearly varies with the evolving dynamics of the pandemic.

\noindent There are different dynamics depending on age groups. Figure \ref{fig:hazard-ages} shows the variations in instantaneous calendar time dependent probability of death given duration in hospital (the time dependent hazard rate)  for patients across different age groups.

The hazard rate estimators described above allow the derivation of further summary measures, such as the median time from admission in hospital to exit (either by recovery or death), conditional on the admission date. Figure \ref{fig:mediantime} presents these median times  for all individuals and stratified by age groups. A prominent peak for people hospitalized by the end of May (consistent across all age groups) illustrates the changing behavior of Covid-19 on distinct calendar times since the onset of the pandemic. Age is an important covariate: whereas changes in the pandemic conditions (variants of the virus and different restrictions regimes) do not have apparently serious impact on the younger population, it seems to strongly affect people in the oldest groups for which the date of hospitalization is crucial.

Variations over time in the durations of stays until recovery or death appear partly driven by changes in the age of patients. In summary, the mean duration of hospitalization for recovered patients has decreased by approximately five days since the first wave, when considering the full sample. The reduction is bigger for people between 60 and 80 years and less significant for ages below 40.
The length of stays has also decreased from about 16 days during the first wave until less than 11 days in the second wave, and about 10 days in the last months of 2021. This is  for the full sample, the reduction being less significant for the younger patients.

Figure \ref{fig:probs} address the question of the probability that a subject who has been hospitalized for $d$ days is discharged alive.
Until March 2021 the probability of getting out alive slightly increases with the hospitalization time, specially for people
above 80 years in the first month in hospital (see a graph of this probability by age groups in  Figure \ref{fig:alive_ages}). In June-2021 this increase seems to revert and we can observe a slight decay. The probability of dying in hospital decreases with hospitalization time for admissions up to June 2021 when it seems to be a slight change of tendency.
During the first and second wave almost 30\% of the older people (above 80) will die in the first day in hospital, this percentage is about 20\% for people between 60 and 80 years, and around 10\% for the younger groups (see Figure \ref{fig:death_ages}). These probabilities are a bit higher in the last months of 2021 which might be explained by the effect of the vaccines and only severe cases going to hospital.

\subsection{Examples of smooth developments of ratio of number of deaths inside versus outside hospital, the Covid-19 case of France}\label{sec:deaths_ratio}
To quantify deaths occurring outside hospitals and thereby estimate the total number of pandemic-related deaths, we track the time evolution of out-of-hospital to in-hospital deaths, denoted by the function $F(t)$ in Section \ref{sec:deaths}. As established in similar studies in survival analysis, see \cite{Nielsen:Tanggaard:01}, that it is more robust to estimate the numerator and the denominator separately and then divide to get the ratio than it is to smooth the ratio directly. 

\begin{figure}[ht]
	\centering
	\includegraphics[width=1\textwidth]{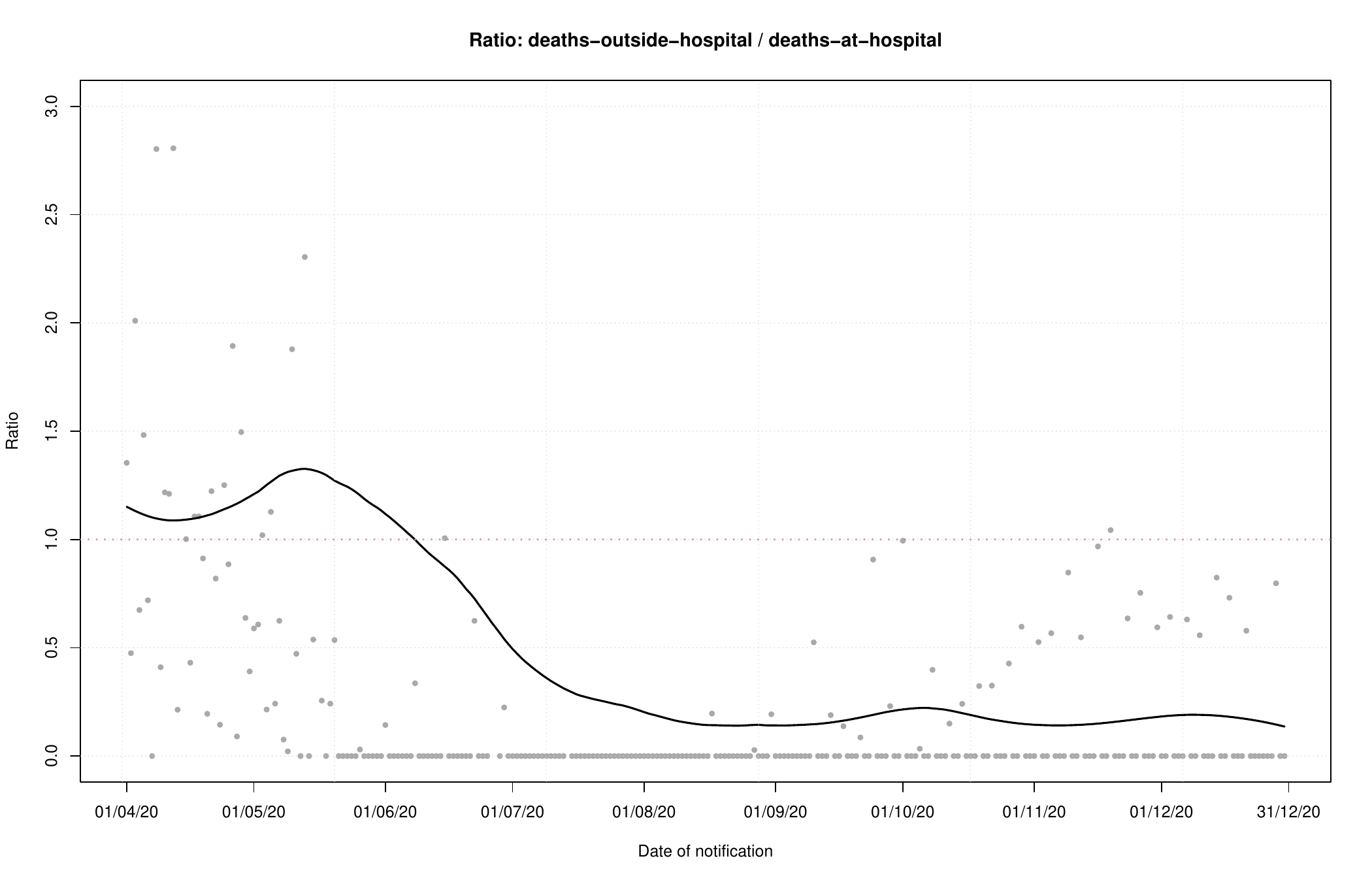}
	\caption{Estimated ratio of number of deaths (inside and outside hospital) from April to December 2020.}
	\label{fig:ratio-dens}
\end{figure}

Figure \ref{fig:ratio-dens} provides the final result of this procedure and we see that, while the ratio has stabilized by the end of the year 2020, with a higher probability for dying in hospital compared to outside hospital, then also this ratio has changed significantly during the evolution of the pandemic during the year 2020. At the onset of the pandemic, more deaths occurred outside hospitals than inside, likely reflecting early challenges in preventing the spread of infection in care homes. The dots in the plot represent the daily ratio between the observed number of deaths outside hospital over the observed number of deaths inside hospital.

\subsection{Examples of extrapolating the ratio of densities with an indicator $C_2$, the Covid-19 case of France } 
For any $t>0$, let  $F(t)$ be the ratio of number of deaths out-of-hospital over the number of deaths in-hospital defined in Section \ref{sec:deaths}, and the estimator $\widehat{F}(t)$ defined as the ratio of the two nonparametric regression estimators, that is $\widehat{F}(t)=\widehat{N}_{\rm out}(t)/\widehat{N}_{\rm in} (t)$, $(0 < t < T)$.

Suppose data are available only up to 30-September-2020 (denoted by $T$). Specifically, we observe the daily number of infections and hospitalizations, deaths inside hospitals, hospital discharges and deaths occurring outside hospitals. The objective is to forecast the total number of deaths (both inside and outside hospital) during the month of October-2020.

To this end, it is first necessary to forecast the number of infected and hospitalized in October-2020. Following the methodology proposed in \cite{Gamiz:etal:25}, we  predict the number of infected in October-2020 using the two-dimensional infection rate estimated from data until 30-September-2020 together with an indicator $C_1$ that closely approximates the well-known reproduction number.

In brief, we assume that the infection rate at the end of the forecasting period equals the more recent estimate multiplied by $C_1$, and varies linearly in between. If no significant changes in the infection are anticipated, setting $C_1=1$ yields reliable predictions for the forthcoming period.

Alternatively, the optimal value of $C_1$ can be estimated in the infeasible scenario in which observations within the forecasting interval are available. In this case, the optimal value is found to be $C_1 = 1.86$ (additional details are provided in \citep{Gamiz:etal:25}). Using this value produces predictions for October 2020 that are closest to the observed infection counts.

Once the predicted number of infections in October is obtained,  the daily number of new hospitalized and deaths inside hospital can be forecasted. This is performed under the assumption that, throughout the forecasting interval, the hospitalization rate (defined in \cite{Gamiz:etal:25}) and the survival hazard function defined (Section \ref{sec:model}) remain constant and equal to the most recent estimates (at time $T$), respectively. 
This procedure yields forecasts of the number of in-hospital deaths for October 2020.

\begin{figure}[ht]
	\centering
	\makebox{\includegraphics[scale=0.4]{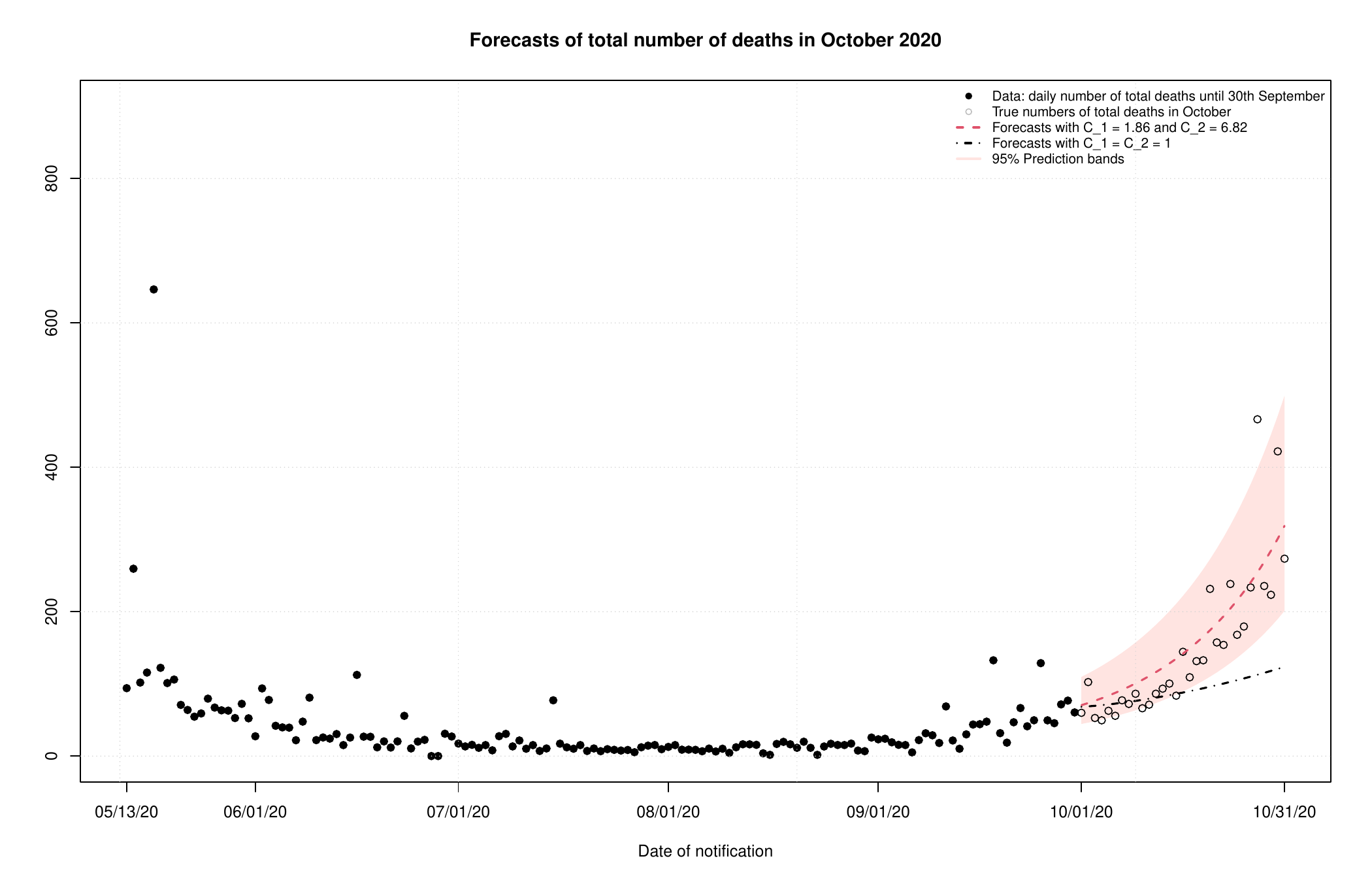}}
	\caption{Total number of deaths from 1-April-2020 to 31-October-2020. The dots in the graph are the observed number of deaths inside hospitals plus observed number of deaths outside hospital until 30-September (black) and in October (grey). Optimal forecasts in October are represented by the blue curve. Forecast obtained with no prior knowledge are represented by the red curve.}\label{fig:forecast}
\end{figure}

Finally, to predict the total number of deaths (inside and outside hospital) we use the extrapolated ratio $\widetilde{F}(T+s)$ as defined in Section \ref{sec:forecast}.
Under the assumption that the ratio of out-of-hospital deaths and in-hospital deaths is constant, that is, $\widetilde{F}(T+s)=\widehat{F}(T)$, for all $0<s \leq h$, we have that $C_2=1$. Alternatively, an infeasible in practice procedure would determine the optimal value of $C_2$ by minimizing the prediction error for daily deaths within the forecasting interval (October2020).\
Figure \ref{fig:forecast} presents the results for the Covid-19 data in France. Two forecasts for the total number of deaths in October are shown. The first (blue line) uses the optimal values $C_1=1.86$ and $C_2=6.82$. The second (red line), assumes no significant changes relative to 30-September-2020  and sets $C_1=C_2=1$.

\section{Discussion}\label{sec:conclusion}
The methodology presented in this paper provides an operational approach to manage a developing pandemic based on easy to communicate and available data. The method can perhaps be understood as a benchmark method to develop in the beginning of a pandemic that can be supplemented during the pandemic by more detailed approaches based on more specific and perhaps more locally defined data. The new management approach to monitor a developing pandemic is able to separate the labour of analysis in the pure statistical forecasting methodology based on available data that can be adjusted according to prior knowledge on the infection reproduction, the $R$-number, and prior knowledge on the distribution of deaths inside and outside hospitals. This potential division of labour might be important when management has to operate fast during (and in particular in the beginning of) a pandemic.

\appendix

\section{Appendix 1. Asymptotic analysis}\label{app:theo1} 
\noindent In this subsection we argue that $ \widehat{\mu} (t/T, t-s)$ obtained in Section \ref{sec:estim3y4} is a consistent  estimator.\\
First, we notice that if the iteration scheme \eqref{eq:mu_r} converges it will converge against the solution $ \widehat{\mu} (t/T, t-s)$  of the equation:
\begin{eqnarray*}
	&&	\widehat{\mu}(t/T, t-s) =\\
	&&	 \frac{{\displaystyle{\int_{0 \leq v < u \leq T}}}{\rm D}_2(s,t,v,u) K_{1,b_1} \left (( t-u)/T\right ) K_{2,b_2} (t-s - (u-v)) \times \frac { \widehat{S}(u/T,v)\widehat{\mu} (u/T, u-v)  N_2(\d v) N(\d u) }{ {\int_0^{u-}}\!\widehat{S}(u/T,w) \widehat{\mu} (u/T, u-w)  N_2(\d w)} }
	{ {\displaystyle{\int_{0 \leq v < u \leq T}}}{\rm D}_2(s,t,v,u) K_{1,b_1} \left ( (t-u)/T\right ) K_{2,b_2} (t-s - (u-v)) \times \frac { \widehat{S}(u/T,v)N_2(\d v) Y(u) \d u} 
		{{ {\int_0^{u-}}}\! \widehat{S}(u/T,w) N_2(\d w)}}. 
\end{eqnarray*}
It can be shown that $\widehat{\mu}(t/T, t-s)$ is a consistent estimator using much the same type of theoretical arguments as \cite{Gamiz:etal:25} uses in the Hawkes process case. The stochastic processes considered in this paper are counting processes and the mathematical statistics used are based on the martingale theory implied by Aalen's multiplicative intensity model, see \cite{Nielsen:98} and \cite{Gamiz:etal:22} for more details. The consistency analysis of this paper can be formulated in this asymptotic framework of standard survival analyses as well with similar derivations and conclusions as in the Hawkes process case discussed in \cite{Gamiz:etal:25}. We omit the theoretical details to save space.

\section{Appendix 2. Simulations}\label{app:simulations}
\begin{figure}[ht]
	\centering
	\includegraphics[width=0.4\textwidth]{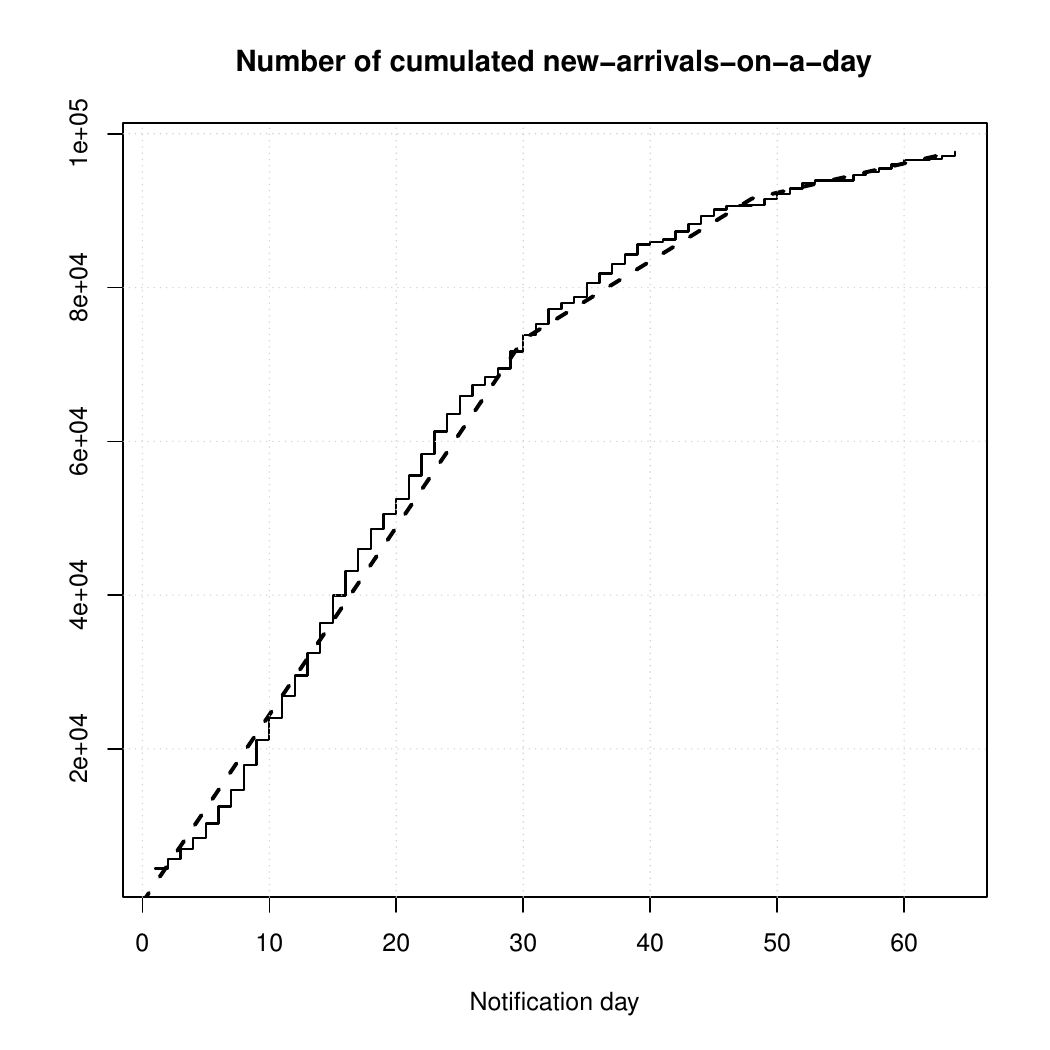} \vskip -.2cm
	\caption{New admissions to hospital are assumed to follow a NHPP. The stepped line are the observed accumulated daily arrivals to hospital and the solid line is the estimated cumulative intensity of the process.}
	\label{fig:newarrivals}
\end{figure}

We carry out a simulation study inspired by the practical application discussed in Section \ref{sec:application}. First of all we assume that new arrivals in  hospital occur according to a non-homogeneous Poisson process whose intensity is assumed piecewise constant and is estimated from data. Figure \ref{fig:newarrivals} displays the observed number of daily accumulated arrivals to hospital and the estimated cumulative intensity curve. 

Let us consider the two-dimensional hazard of survival time in hospital as $\mu(t,w)=\alpha(t) \alpha(w)$. Here $t$ refers to the notification date and $w$ is  time duration in hospital.

We denote $\mu(t,w)$ the two-dimensional hazard for time duration in hospital regardless the final event is death or recovery, then $\mu(t,w)=\mu_1(t,w)+\mu_2(t,w)$, where $\mu_1(t,w)=\alpha(t)\alpha_1(w)$ is the two-dimensional hazard of survival time in hospital until death and $\mu_2(t,w)=\alpha(t)\alpha_2(w)$ is the hazard of survival time until recovery.

\begin{figure}[ht]
	\centering
	\includegraphics[width=0.95\textwidth]{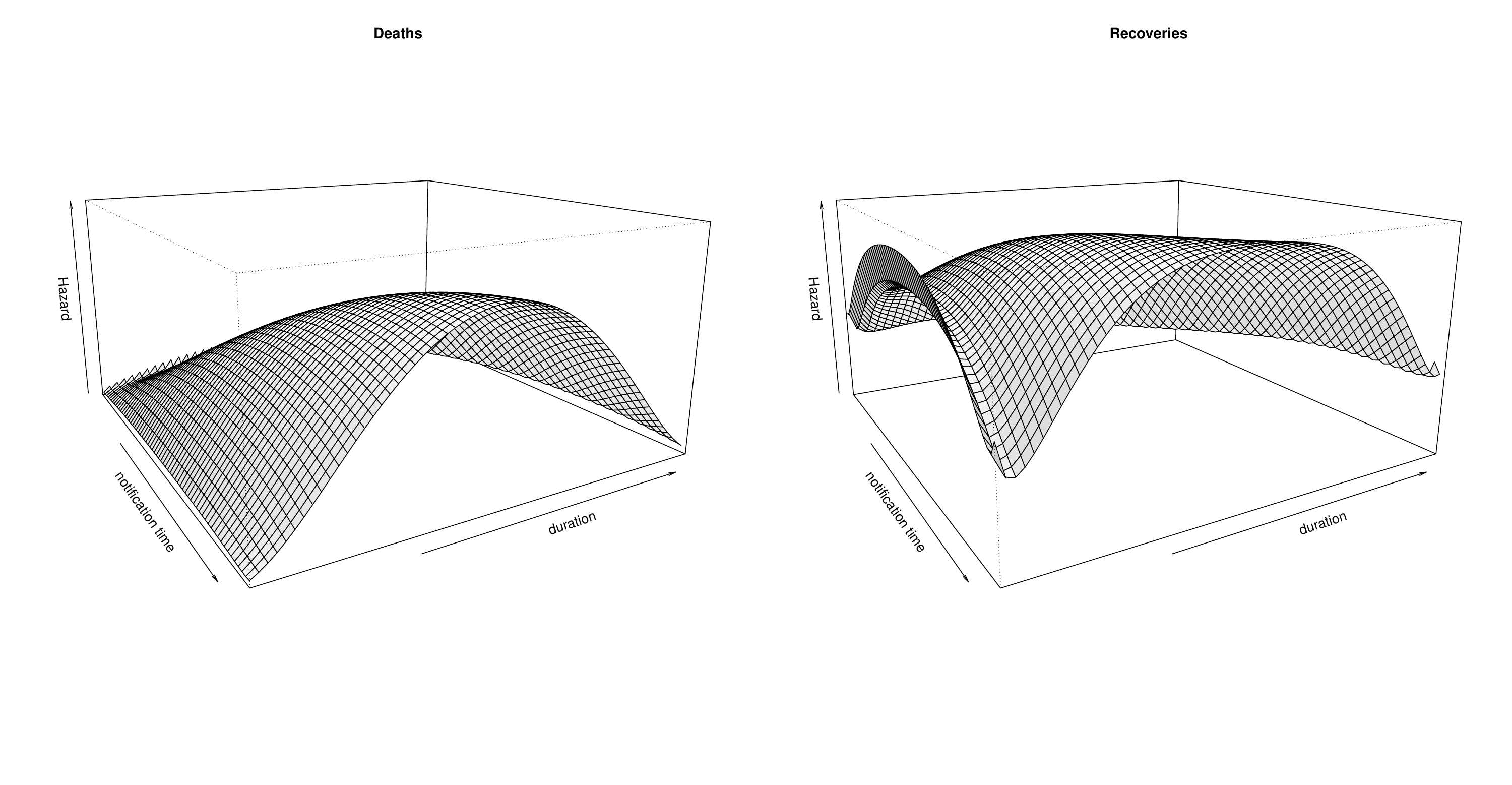}
	\caption{True models to simulate data. Left panel:  hazard for survival time in hospital until death, $\mu_1(t,w)$. Right panel:  hazard for survival time in hospital until recovery, $\mu_2(t,w)$.}
	\label{fig:truemodels}
\end{figure}

The  true hazard rates are:
\begin{align}
	\alpha_1(t) &= B(t;2,2)/T,  \nonumber \\ 
	\alpha_2(t) &= (0.6/T)\left\{ B(t;0.5,0.5)+B(t;2,4)+B(t;4,2)\right\}, \nonumber \\ 
	\alpha(t) &=\alpha_1(t)+\alpha_2(t), \nonumber
\end{align}
where $t \in (0, T)$ and $B(t;a,b)$ is the density at $t$ of a Beta distribution with parameters $(a,b)$. See Figure \ref{fig:truemodels}.

To evaluate the performance of the estimators, we have considered the following measure of the error
\[
\text{err}(\widehat{\mu}_{j,{\bf {\rm b}}})=n^{-1} \sum_{t=1}^M \sum_{w=1}^{t} \left[\widehat{\mu}_{j,\bf {\rm b}}(t,w)-\mu_{j}(t,w)\right]^2
\]
with $j=1,2$.\\
The results of the simulations are presented in Table \ref{tab:3}. Two estimates are compared: the local linear hazard estimate using full information and the hazard estimates derived from our algorithm using available data.  
We can see that using partial information notably increases the integrated bias and thus the $MISE$.

\begin{table}[H]
	\caption{Simulation results based on 500 simulations. For each case (method and sample size) the average of the $ISE$ values ($MISE$), are decomposed into squared integrated bias ($ISB$) and integrated variance ($MIV$). Numbers have been multiplied by $10^9$. }
	\label{tab:3}
	\centering
	\begin{tabular}{ccllll}
	\hline
	\multicolumn{2}{c}{} & \multicolumn{2}{c}{Full information} & \multicolumn{2}{c}{Partial information} \\ \hline %
	N & Criteria & Deaths & Recoveries & Deaths & Recoveries \\ \hline %
	10000 & MISE & 1.6755 & 2.8311 & 11.7095 & 9.6399 \\
	& ISB & 0.6271 & 1.1515 & 9.2786 & 4.3358 \\ 
	& MIV & 1.0484 & 1.6797 & 2.4309 & 5.304 \\ 
	1e+05 & MISE & 0.0394 & 0.0978 & 0.7396 & 0.3766 \\  
	& ISB & 0.0154 & 0.0601 & 0.6926 & 0.2612 \\ 
	& MIV & 0.024 & 0.0376 & 0.047 & 0.1155 \\ 
	1e+06 & MISE & 0.001 & 0.0039 & 0.0699 & 0.0472 \\  
	& ISB & 4e-04 & 0.0022 & 0.0698 & 0.047 \\ 
	& MIV & 5e-04 & 0.0017 & 1e-04 & 2e-04 \\ 
	\hline
\end{tabular}
\end{table}

\section{Appendix 3. Data adjustment for variations by day of the week} \label{app:koyama} 
The analysis of Covid-19 pandemic data presented in this paper reveals consistent variations in reported cases throughout the week, with notably fewer cases typically recorded on weekends. To correct for this effect we apply a weekday adjustment similar to the method suggested in \cite{Koyama:et:al:21}. Given a sequence of daily case counts $\{N_i;i=1,\ldots,T\}$, where $T$ is the total number of days, we define a mapping from each day $i \in \{1,\ldots, T\}$ to a weekday index, i.e. $ d(i) \in \{1, \ldots, 7\}$, with Monday as 1 and Sunday as 7 (or any consistent 7-day indexing), so that, for example, if day $i$ falls on a Monday, then $d(i) = 1$, and so on.\\
For each day of the week, we compute the total number of reported cases
\[
f_d = \sum_{\substack{i = 1 \\ d(i) = d}}^T N_i,\quad \text{for } \  d \in \{1, \ldots, 7\}.
\]
Let $n = \sum_{i=1}^T N_i$ denote the total number of reported cases across all days. The weight for day $i$ is then defined as
\[
w_{d(i)} = \frac{f_{d(i)}}{n/7}, \quad \text{for }   i=1,\ldots, T.
\]
These weights represent the relative contribution of each weekday to the total count, scaled such that their average is 1, that is $(1/7)\sum_{d} w_d=1$. If no weekday effect is present, all weights $w_d$ would be close to 1.\\
Finally, the adjusted daily counts $\{\tilde{N}_i, i=1,\ldots, T\}$ are obtained by dividing each original count by its corresponding weekday weight
\[
\tilde{N}_i = {N_i}/{w_{d(i)}} \quad \text{for } i = 1, \ldots, T.
\]
This procedure reduces the influence of systematic underreporting or overreporting on specific weekdays, making the adjusted series more suitable for time-series analysis and modelling.

\section{Appendix 4. Constructing Confidence and Prediction Bands via Bootstrap} \label{app:bootstrap}

Let $\gamma$ denote the estimated factor of variance of the process defined in \eqref{eq:gamma} based on observations $N_{1,0}, N_{1,1}, \ldots, N_{1,T}$ of the infection process. 
\newpage

\begin{algorithm}[H]
	\caption{ \textbf{Part I: Bootstrap generation}}
	\label{alg:uq1}
	\KwIn{Parameter $\gamma$; initial infected $N_{1,0}$; estimations $\hat{\mu}_1, \hat{\mu}_2, \hat{\mu}_3, \hat{\mu}_4$; size of the observation window $T$; number of bootstrap samples $R$.}
	\KwOut{$R$ bootstrap trajectories of size $T$ of the processes $N_1, N_2, N_3, N_4$.}
	
	\tcc{Step 1: User selects $k > \gamma$}
	Take $k>\gamma$\;
	
	\BlankLine
	\tcc{Step 2: Parameter setup}
	Set $\beta = (\gamma-1)/(k(k-1))$ and $\alpha = 1-k\beta$\;
	\BlankLine		
	\For{$r \leftarrow 1$ \KwTo $R$}{
		\tcc{Step 3: Generate $N_{1,i}^{(r)}$}
		Take $N_{1,0}^{(r)}:=N_{1,0}$\;
		\For{$i \leftarrow 1$ \KwTo $T$}{
			Compute 
			\[
			\lambda^{(r)}_{1,i} = N^{(r)}_{1,1}\hat{\mu}_1(i/T,i) + N^{(r)}_{1,2}\hat{\mu}_1(i/T,i-1) + \cdots + N^{(r)}_{1,i-1}\hat{\mu}_1(i/T,1)
			\]
			Generate 
			\[
			N^{(r)}_{1,\alpha,i} \sim \mathrm{Pois}(\alpha \lambda^{(r)}_{1,i}), \quad 
			N^{(r)}_{1,\beta,i} \sim \mathrm{Pois}(\beta \lambda^{(r)}_{1,i})
			\]
			Set 
			\[
			N^{(r)}_{1,i} = N^{(r)}_{1,\alpha,i} + k\,N^{(r)}_{1,\beta,i}
			\]
		}
		
		\BlankLine
		\tcc{Step 4: Generate $N_{2,i}^{(r)}$}
		\For{$i \leftarrow 1$ \KwTo $T$}{
			Compute 
			\[
			\lambda^{(r)}_{2,i} = N^{(r)}_{1,1}\hat{\mu}_2(i/T,i) + \cdots + N^{(r)}_{1,i}\hat{\mu}_2(i/T,1)
			\]
			Generate $N^{(r)}_{2,i} \sim \mathrm{Pois}(\lambda^{(r)}_{2,i})$
		}
		
		\BlankLine
		\tcc{Step 5: Generate $N_{3,i}^{(r)}$ and $N_{4,i}^{(r)}$}
		\For{$i \leftarrow 1$ \KwTo $T$}{
			\For{$d \leftarrow 1$ \KwTo $i$}{
				Compute 
				\[
				\lambda^{(r)}_{3,i}(d) = N^{(r)}_{2,i-d+1}\hat{S}(i/T,i-d+1)\hat{\mu}_3(i/T,d), \quad
				\lambda^{(r)}_{4,i}(d) = N^{(r)}_{2,i-d+1}\hat{S}(i/T,i-d+1)\hat{\mu}_4(i/T,d)
				\]
				with 
				\[
				\hat{\mu} = \hat{\mu}_3 + \hat{\mu}_4, \quad
				\hat{S}(i/T,i-d+1) = \prod_{j=1}^{d-1}[1-\hat{\mu}((i-j+1)/T,j)]
				\]
				Generate 
				\[
				N^{(r)}_{3,i,d} \sim \mathrm{Pois}(\lambda^{(r)}_{3,i}(d)), \quad 
				N^{(r)}_{4,i,d} \sim \mathrm{Pois}(\lambda^{(r)}_{4,i}(d))
				\]
			}
			Obtain
			\[
			N^{(r)}_{3,i} = \sum_{d=1}^i N^{(r)}_{3,i,d}, \quad 
			N^{(r)}_{4,i} = \sum_{d=1}^i N^{(r)}_{4,i,d}
			\]
		}
	}
\end{algorithm}

\addtocounter{algocf}{-1} 
\begin{algorithm}[H]
	\caption{\textbf{Part II: Forecasting and adjustment}}
	\label{alg:uq2}
	\KwIn{Bootstrap samples $N^*_{1,i}, N^*_{2,i}, N^*_{3,i}, N^*_{{\rm in},i}$,  ($i=0,\ldots, T$) from Part I; size of the prediction window $h$; constants $C_1, C_2$; and estimated ratio ${\hat F}(T)$.}
	\KwOut{Predicted number of occurrences $N^*_{1,T+i}, N^*_{2,T+i}, N^*_{3,T+i}, N^*_{{\rm in},T+i}$; adjusted total number of deaths  $N^*_{{\rm d},T+i}$  ($i=1,\ldots,h$).}
	
	\For{$r \leftarrow 1$ \KwTo $R$}{
		\tcc{Step 1: Estimate $\hat{\mu}_1^{(r)}$, $\hat{\mu}_2^{(r)}$, $\hat{\mu}_3^{(r)}$, $\hat{\mu}_4^{(r)}$.}
		
		\BlankLine
		\tcc{Step 2: Predict $N_{1,T+i}^{(r)}$}
		\For{$i \leftarrow 1$ \KwTo $h$}{
			Compute 
			\[
			\widetilde{\mu}_1^{(r)}(T+i,d) = \hat{\mu}_1^{(r)}(i/T,d)\{1+(C_1-1)i/h\}, \quad d=1,\ldots,T
			\]
			
			Obtain 
			\[
			N^{(r)}_{1,T+i} = \sum_{j=1}^{T} N^{(r)}_{1,i+j-1}\,\widetilde{\mu}_1^{(r)}(T+i,T-j+1)
			\]
		}
		\BlankLine
		\tcc{Step 3: Predict $N_{2,T+i}^{(r)}$}
		\For{$i \leftarrow 1$ \KwTo $h$}{
			Compute 
			\[
			\widetilde{\mu}_2^{(r)}(T+i,d) = \hat{\mu}_2^{(r)}(i/T,d), \quad d=1,\ldots,T
			\]
			Obtain 
			\[
			N^{(r)}_{2,T+i} = \sum_{j=1}^{T} N^{(r)}_{1,i+j}\,\widetilde{\mu}_2^{(r)}(T+i,T-j+1)
			\]
		}
		
		\BlankLine
		\tcc{Step 4: Predict $N_{{\rm in},T+i}^{(r)}$}
		\For{$i \leftarrow 1$ \KwTo $h$}{
			Compute 
			\[
			\widetilde{\mu}_3^{(r)}(T+i,d) = \hat{\mu}_3^{(r)}(i/T,d),  \quad d=1,\ldots,T
			\]
			and
			\[
			\widetilde{\mu}_4^{(r)}(T+i,d) = \hat{\mu}_4^{(r)}(i/T,d),  \quad d=1,\ldots,T
			\]
			and
			\[
			\widetilde{S}(i,i-d+1) = \prod_{j=1}^{d-1}[1-(\widetilde{\mu}_3^{(r)}+\widetilde{\mu}_4^{(r)})(i-j+1,j)]
			\]		
			Obtain 
			\[
			N^{(r)}_{{\rm in},T+i} = \sum_{j=1}^{T} N^{(r)}_{2,i+j}\,\widetilde{S}(i,i-j+1)\,\widetilde{\mu}_4(T+i,T-j+1)
			\]
		}
		
		\BlankLine
		\tcc{Step 5: Final adjustment}
		\For{$i \leftarrow 1$ \KwTo $h$}{
			Compute
			\[
			N^{(r)}_{{\rm d},T+i} =N^{(r)}_{{\rm in},T+i} \left[1+\hat{F}(T)\left\{1+(C_2-1)i/h\right\}\right]
			\]
		}
	}
\end{algorithm}

Compute  2.5\% and 97.5\% quantiles of the bootstrap samples at each day $i+T (i=1,\ldots, h)$ in the forecasting horizon, for the  total number of deaths inside and outside the hospital (see Figure \ref{fig:forecast}).


\end{document}